\documentclass[aps,twocolumn,nofootinbib,amsmath,amssymb,a4paper,showkeys,notitlepage,byrevtex,longbibliography]{revtex4-2}

\def\pdftitle{MagOD}
\def\authorname{Leon Abelmann}
\def\pdfsubject{Performance and application of a simple automated magnetic
optical density meter for analyzing magnetotactic bacteria}
\def\pdfkeywords{}
\def\pdfbackref{none}

\usepackage[
        colorlinks=true,
        urlcolor=darkblue,               
        anchorcolor=darkblue,
        filecolor=green,                   
        linkcolor=darkblue,              
        menucolor=darkblue,
        citecolor=darkblue,
        pagebackref,
        backref={\pdfbackref},
        pdfpagemode={UseOutlines},
        bookmarks=true,
        bookmarksopen=true,
        pdftitle={\pdftitle},
        pdfauthor={\authorname}, 
        pdfsubject={\pdfsubject},
        pdfkeywords={\pdfkeywords}
        ]{hyperref}
\usepackage{thumbpdf}

\usepackage{graphicx}
\DeclareGraphicsExtensions{.pdf,.png,.jpg}
\graphicspath{{~/Documents/Images//}{Figures//}}

\usepackage{color}
\definecolor{darkgreen}{rgb}{0,0.5,0}
\definecolor{darkblue}{rgb}{0,0,0.5}
\definecolor{brown}{rgb}{0.98,0.92,0.73}
\definecolor{red}{rgb}{1,0,0}
\definecolor{yellow}{rgb}{1,1,0}
\definecolor{blue}{rgb}{0,0,1}
\definecolor{green}{rgb}{0,1,0}
\definecolor{purple}{rgb}{1,0,1}
\definecolor{gray}{rgb}{0.8,0.8,0.8}
\definecolor{black}{rgb}{0,0,0}
\definecolor{white}{rgb}{1,1,1}
\definecolor{gold}{rgb}{1.,0.84,0.}

\usepackage{amsmath}
\usepackage{wasysym}
\usepackage{nicefrac}

\usepackage{booktabs}

\usepackage{lineno}

\usepackage{pdfsync}

\usepackage[mode=math]{siunitx}

\usepackage{booktabs}
\usepackage{tabularx}
\usepackage{dcolumn}

\newcommand{\CC}{C\nolinebreak\hspace{-.05em}\raisebox{.4ex}{\tiny\bf +}\nolinebreak\hspace{-.10em}\raisebox{.4ex}{\tiny\bf +}}
\def\CC{{C\nolinebreak[4]\hspace{-.05em}\raisebox{.4ex}{\tiny\bf ++}}}

\def\smallfigurewidth{0.6\columnwidth}
\def\figurewidth{0.8\columnwidth}
\def\widefigurewidth{\columnwidth}

\newif\ifcmtr
\cmtrfalse
\ifcmtr 
\newcommand{\cmtr}[1]{ %
   [\color{red} \textbf{#1} \normalcolor]%
}%
\else
\newcommand{\cmtr}[1]{%
}%
\fi

\newif\ifcmtrj
\cmtrjfalse
\ifcmtrj 
\newcommand{\cmtrj}[1]{ %
   [\color{green} \textbf{#1} \normalcolor]%
}%
\else
\newcommand{\cmtrj}[1]{ %
}%
\fi

\begin{document}
\title{An open-source automated magnetic optical density meter for analysis of suspensions of magnetic cells and particles}

\date{\today}
\author{Marcel K. Welleweerd$^{1}$}
\author{Tijmen Hageman$^{1,2}$}
\author{Marc Pichel$^{1,2}$}
\author{Dave van As$^{1}$}
\author{Hans Keizer$^{1}$}
\author{Jordi Hendrix$^{1}$}
\author{Mina M. Micheal$^{1}$}
\author{Islam S.M. Khalil$^{1}$}
\author{Alveena Mir$^{2}$}
\author{Nuriye Korkmaz$^{2}$}
\author{Robbert Kr\"awinkel$^{1}$}
\author{Daniel M. Chevrier$^{3}$}
\author{Damien Faivre$^{3}$}
\author{Alfred Fernandez-Castane$^{4}$}
\author{Daniel Pfeiffer$^5$}
\author{Leon Abelmann$^{1,2}$}
\affiliation{
  $^1$University of Twente, EWI/Robotics and Mechatronics, PoBox 217, 7500 AE Enschede, The Netherlands
  $^2$KIST Europe, Biosensors Group, Campus E7, 66123 Saarbr\"ucken, Germany,
  $^3$Aix-Marseille Universit\'e, CEA, CNRS, BIAM, UMR7265,
  13108  Saint-Paul lez Durance, France,
  $^4$Energy and Bioproducts Research Institute, Aston University,
  Birmingham, B4 7ET, UK
  $^5$Lehrstuhl f\"ur Mikrobiologie, Universit\"at Bayreuth,
  Universit\"atsstrasse 30,
  95447 Bayreuth\\
  The author to whom correspondence may be addressed: \underline{l.abelmann@kist-europe.de}}

\begin{abstract}
  We present a spectrophotometer (optical density meter) combined with
  electromagnets dedicated to the analysis of suspensions of
  magnetotactic bacteria. The instrument can also be applied to
  suspensions of other magnetic cells and magnetic particles. We have
  ensured that our system, called MagOD, can be easily reproduced by
  providing the source of the 3D prints for the housing, electronic
  designs, circuit board layouts, and microcontroller software. We
  compare the performance of our system to existing adapted
  commercial spectrophotometers. In addition, we demonstrate its use
  by analyzing the absorbance of magnetotactic bacteria as a function
  of their orientation with respect to the light path and their speed
  of reorientation after the field has been rotated by
  \SI{90}{\degree}. We continuously monitored the development of a
  culture of magnetotactic bacteria over a period of five days, and
  measured the development of their velocity distribution over a
  period of one hour. Even though this dedicated spectrophotometer is
  relatively simple to construct and cost-effective, a range of
  magnetic field-dependent parameters can be extracted from
  suspensions of magnetotactic bacteria. Therefore, this instrument
  will help the magnetotactic research community to understand and
  apply this intriguing micro-organism.
\end{abstract}

\maketitle 

\tableofcontents


\section{Introduction}

Magnetotactic bacteria biomineralize a chain of iron-oxide or
iron-sulfide nanocrystals (a magnetosome) that makes them align with
the Earth's magnetic field~\cite{Frankel1979, Farina1990}. This
property allows them to search efficiently for the optimal redox
conditions in stratified water columns~\cite{Lefevre2014}. Ever since
their discovery~\cite{Bellini2009,Blakemore1979} they have intrigued
researchers in magnetism, not in the least because one can easily
control them in a
microscope~\cite{Lee2004b,Khalil2013b,Pichel2018}. Magnetotactic
bacteria are used as model systems for many applications of magnetic
particles, such as magnetic domain
imaging~\cite{Futschik1989,Harasko1995},
hyperthermia~\cite{Hergt2005,Gandia2019}, magnetic particle
imaging~\cite{Makela2022}, microrobotic
manipulation~\cite{Martel2010}, targeted drug
delivery~\cite{Martel2009,Khalil2013,DeLanauze2014,Felfoul2016} and
studies of spin-wave propagation~\cite{Zingsem2019}.

Rosenblatt~\cite{Rosenblatt1982} discovered that
the transmission of light through suspensions of magnetotactic
bacteria is influenced by the direction of an externally applied
field.  This effect has been successfully applied as
a simple method to monitor such processes as the cultivation of
magnetotactic bacteria~\cite{Faivre2010,
  Fernandez2018,Yang2013a,Song2014} and to assess their
velocity~\cite{Lefevre2009,Chen2014}.

\subsection{Research question and relevance}
Commonly, the field-dependent transmission of light through a
suspension of  magnetotactic bacteria is measured by extending a standard spectrophotometer with a magnetic add-on. Such
spectrophotometers are also known as optical density meters, and are
commonly used in biolabs to determine cell concentrations.

The modification of existing spectrophotometers with magnetic add-ons
has several disadvantages. (i)~These instruments are relatively
complex and expensive, so modifications are usually made to depreciated
equipment. (ii)~Most instruments contain magnetic components that disturb
the field and there is generally little space to mount electromagnets,
certainly not in three dimensions. (iii)~The various types of
spectrophotometers and magnetic field generators and the variations
between laboratories lead to a lack of a standardized measurement. (iv)~More fundamentally, most spectrometers are not intended for sub-second
continuous registration of absorbance over time. They are operated
manually, and often use flash lamps.

\begin{figure}
  \centering
  \includegraphics[width=\widefigurewidth]{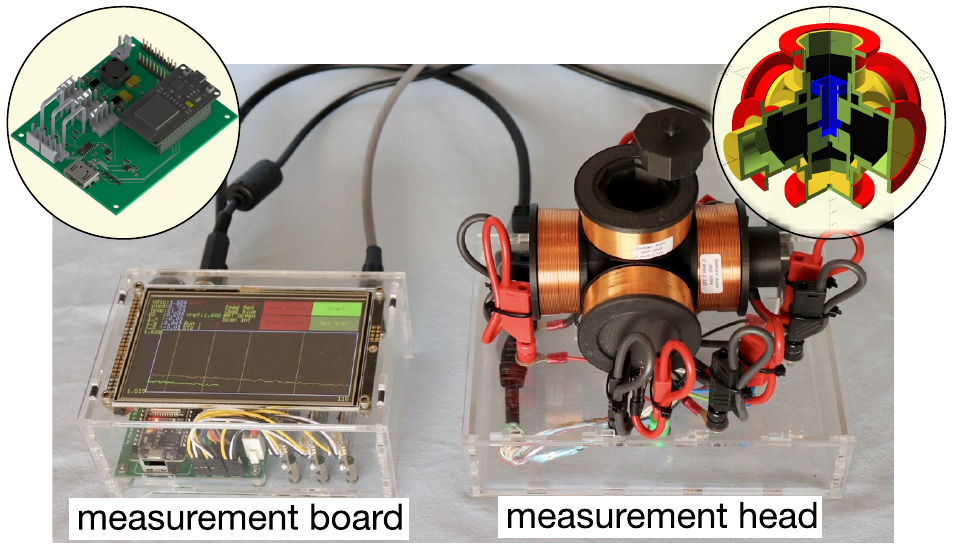}
  \caption{Photograph of an open-source spectrophotometer with
    magnetic field option (MagOD). The system consists of a measurement
    head (right) in which a cuvette with a suspension of magnetotactic
    bacteria is inserted. The measurement board (left) is dedicated
    to controlling the magnetic field, data acquisition and
    communication with the user over a touchscreen and WiFi. The
    design of the system is open, including the layout of the electronic
    circuit boards (top left), 3D print source files (top right) and
    control software.}
  \label{fig:MagOD}
\end{figure}

In this publication, we present a spectrophotometer that intimately
integrates the optical components with a magnetic field system and is
dedicated to research on magnetotactic bacteria, see Figure~\ref{fig:MagOD}. Additionally, the design considers that students
at the Master's or early PhD level should be capable of constructing such
an instrument, both with respect to complexity and cost. Our main
research question addresses how this new magnetic optical density meter, which we have dubbed MagOD, compares to existing adapted spectrophotometers, and what
novel measurement strategies it enables.

\subsection{Previous work}
The system  to be  constructed is  a spectrophotometer combined with a magnetic field system. It is therefore useful to
compare it with commercial spectrophotometers. These systems generally
use a xenon light source and monochromator with a wide wavelength
range. Table~\ref{tab:meters} provides an overview of the specifications of
representative commercial systems (Biochrome Ultrospecs and the
Eppendorf Biophotometer used here for comparison), including their
wavelength range $\lambda_\text{min} - \lambda_\text{max}$,
spectral bandwidth $\Delta_\lambda$, maximum absorbance
$O_\text{D}$ (see Eq.~(\ref{eq:OD})) and approximate price.

The first spectrophotometer modified with a magnetic field module was
presented by Sch\"uler~\cite{Schuler1995}. That device was based on standard optical components and used a permanent magnet to generate a \SI{70}{mT} field. Later versions were constructed around commercial
optical density meters such as the ones presented by Lef\`evre \cite{Lefevre2009}
(based on a Varian Cary 50 UV) and Song~\cite{Song2014} (based on a
Hitachi U2800). In their case, the magnetic field is generated by coil
systems that can generate adjustable fields up to \SI{6}{mT}~\cite{Lefevre2009}.

Table~\ref{tab:meters} also lists the parameters of the MagOD system  introduced in this paper. Its optical properties and price range compare well with those of standard commercial
systems, whereas its field range is similar to that of the adapted systems by
Lef\`evre and Song.

\cmidrulewidth=.03em
\renewcommand{\arraystretch}{1.1} 
\begin{table}
  \caption{Optical density meters.}
  \label{tab:meters}
  \begin{ruledtabular} 
    \begin{tabular}{@{} l cccccc @{} }
      & $\lambda_\text{min}$ & $\lambda_\text{max}$ & $\Delta_\lambda$ & $O_\text{D}$    & $B$ & Price\\
      & (nm) & (nm) & (nm) & & (mT) & (Eu)\\
      \cmidrule(lr){2-7} 
      Ultrospec 8000 & 190 & 1100 & 0.5&8& &12000\\
      Biophotometer D30&230&600&4&3& &5000\\
      Ultrospec 10&600&600&40&2.3& &1300\\
      \\
      Sch\"uler~\cite{Schuler1995}, 1995&637&637&18&&70&\\
      Lef\`evre~\cite{Lefevre2009}, 2009&190&1100&1.5&3.3&0-6&\\
      Song~\cite{Song2014}, 2014&190&1100&1.5&6&0-4.3&\\
      MagOD&465&640&25&2&0-5&2000\\

     \end{tabular}
   \end{ruledtabular}
 \end{table}

\subsection{Structure and contents}
In this paper, we first discuss a model of the relationship between
the transmission of light and the orientation of magnetotactic
bacteria (Section~\ref{sec:theory}).  Next to the specifications
listed in Table~\ref{tab:meters}, we defined other specifications that
are important for  analyzing magnetotactic bacteria and the open-source nature of the instrument. Our design choices are discussed in
Section~\ref{sec:design}.  The results section is divided into two
parts. In Section~\ref{sec:performance}, we analyze the performance of
our current implementation and compare it with a commercial optical
density meter. Section~\ref{sec:applications} illustrates the
possibilities of our novel system by giving four examples of
experiments to extract information about the magnetic behavior of
magnetotactic bacteria. This instrument is still very much a work in
progress, and we invite the magnetotactic bacteria community to
participate in its further development. For this purpose, we indicate possibilities for improvement and
ideas for additional applications in Section~\ref{sec:discussion}.


\section{Theory}
\label{sec:theory}

The standard method to determine the proportion of bacteria with
magnetosomes in a culture is to observe the changes of light
transmitted through a suspension of bacteria under rotation of a
magnetic field. This technique was pioneered by
Rosenblatt~\cite{Rosenblatt1982}. The transmission of light is
dependent on the relative orientation of the bacteria to the light
path. For MSR-1, which are long, slender bacteria, transmission is
high when the field is perpendicular to the light path, whereas it is
low when the field is aligned parallel to the light path. This is
somewhat counterintuitive, as MSR-1 have the smallest projected cross
section when they are aligned along the line of view. (So in contrast
to blinds, MSR-1 let light pass if the blinds are closed).

It is important to realize that we measure the intensity of light
that reaches the photodetector. The light leaving the  source can
either be absorbed by the suspension of bacteria or  scattered
sideways so that it does not reach the photodetector. Highly dense
suspensions of magnetotactic bacteria are milky white in appearance. Like milk, it is therefore very likely that magnetotactic
bacteria scatter, rather than absorb, light. MSR-1 are small compared
to the wavelength of the incident light, especially considering the size of their
cross section. Additionally, their refraction index is only
slightly higher than that of the surrounding liquid. These small ``optically
soft'' objects scatter more light in forward direction if their
projected area along the light path increases~\cite{Bennet2016}. This
would explain why the light intensity at the photodetector drops if
the MSR-1 are aligned with the light beam.

For MSR-1, the projected area is roughly proportional to the sine of
the angle between the long axis of the bacteria body and the light
path. Owing to Brownian motion and flagellar movement, the bacteria will
not be aligned perfectly along the field direction but show an angular
distribution. The width of this distribution will decrease with
increasing field. In the following discussion, we develop a simple theory to
account for this effect. As the MagOD meter allows us to 
adjust the angle and strength of the magnetic field accurately, we can use it to
validate the approximation.

\subsection{Angle-dependent scattering $C_\text{mag}$}
\begin{figure}
  \centering
    \includegraphics[width=\smallfigurewidth]{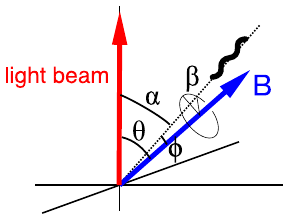}
    \caption{Definitions of various angles. In our MagOD system, we set the
      angle $\theta$ between the light path and the magnetic field
      $B$. The bacteria align with the direction of the field, but can
      deviate by a small angle $\phi$ in a cone around the field
      direction described by $\beta$. As a result, the angle between
      the bacteria long axis and the light is $\alpha$. In case of
      sufficiently large fields, $\phi=0$, $\alpha=\theta$, and $\beta$
    is irrelevant.}
    \label{fig:angledefinitions}
\end{figure}

We define the angle between the light path and the
MSR-1 long axis as $\alpha$,  see
Figure~\ref{fig:angledefinitions}, and introduce a scattering factor
relative to the intensity of light reaching the photodetector
($I(\alpha)$ with unit \si{V}) 
\begin{equation}
  g(\theta)=\frac{I_\text{max}-I(\alpha)}{I_\text{max}-I_\text{min}} \, .
  \label{eq:gdef}
\end{equation}
For MSR-1, the photodetector signal $I$ has a maximum when the MSR-1
are aligned perpendicular to the light beam
($I_\text{max} = I(90) = I_\bot$), at which point scattering $g(90)$
is minimal.

Sch\"uler~\cite{Schuler1995} introduced a parameter to characterize
the relative proportion of magnetotactic bacteria by comparing the
light reaching the detector for the magnetic field aligned parallel
and perpendicular to the light path ($C_\text{mag}$ denotes the ``coefficient of
magnetically induced differential light scattering'' or the ``ratio of
scattering intensities''~\cite{Katzmann2013}). Assuming that the
scattering intensity can be estimated from the reduction of light
reaching the detector  compared to the reference value of a sample
without bacteria ($I_\text{ref}$), the original definition is
\begin{equation}
	C^*_{\mathrm{mag}} =
        \frac{I_\text{ref}-I(0)}{I_\text{ref}-I(90)} \, . \nonumber
\end{equation}

With increasing concentration of bacteria, the total amount of light
reaching the photodetector will decrease. In microbiology,  cultures are traditionally characterized by ``optical density'', a parameter that relates the reduction in light
intensity to the reference value on a \num{10}-base log
scale\footnote{Analogous to the Beer--Lambert law. However, it should be noted that the relation between $O_\text{D}$ and cell
  concentration is only approximate~\cite{Myers2013}.}
\begin{equation}
  \label{eq:OD}
  O_\text{D}(\alpha) = \log\left(\frac{I_\text{ref}}{I(\alpha)}\right) =
  \log\left(I_\text{ref}\right)-\log\left(I(\alpha)\right) \, .
\end{equation}
After the pioneering work of Sch\"uler, researchers started to equip
these optical density meters with magnetic fields~\cite{Song2014,Katzmann2013,Zhao2007,Lang2008}. Using
these instruments, it is more convenient to define $C_\text{mag}$
as
\begin{equation}
  \label{eq:Cmag}
	C_\text{mag} =
        \frac{O_{\text{D}\parallel}}{O_{\text{D}\bot}}
        =\frac{\log(I_\text{ref})-\log(I(0))}{\log(I_\text{ref})-\log(I(90))} \, .
\end{equation}
Today, the latter definition is commonly used. However, it should be noted that the values are not identical, not even for $C_\text{mag}$
close to unity (see Appendix~\ref{sec:appendix_cmag}). As $C_\text{mag}$ equals unity  in the absence of
magnetotactic bacteria, 
($C_\text{mag}-1$) is often
plotted~\cite{Schuler1998,Katzmann2013,Heyen2003, Yang2013a, Lang2008}.

In addition to the ratio, it is insightful to study the absolute difference
between the absorbances in the parallel  and perpendicular directions
\begin{align}
  \label{eq:deltaOD}
  \Delta_\text{OD} &= O_{\text{D}\parallel} - O_{\text{D}\bot} \nonumber\\
  &= \log(I(90))-\log(I(0)) \, .
\end{align}
This difference is proportional to the absolute amount of
magnetotactic bacteria that rotate in the field.

\subsection{Dynamic response}
When measuring $C_\text{mag}$ with adapted photospectrometers, the
$O_\text{D}$ values are measured over a long interval, whereas the
actual rotation of the bacteria is not measured. However, our MagOD
system can measure at sub-second intervals and monitor the dynamic
behavior of the bacteria.  The response of bacteria to a change in
field direction is determined by the balance between magnetic torque
and rotational drag
torque~\cite{Esquivel1986,Steinberger1994,Zahn2017,Pichel2018}. Alignment
of a bacterium to an external magnetic field with angle $\phi(t)$ (see
Figure~\ref{fig:angledefinitions}) can be described by a simple
differential equation
\begin{equation}
  f \frac{\partial\phi(t)}{\partial t} + m B \sin\phi(t)= 0 \, , \nonumber
\end{equation}
where $f$ [\si{\newton\meter\second}] represents the rotational drag coefficient, $m$ 
[\si{\ampere\square\meter}] the magnetic dipole moment of the bacterium, and $B$ 
[\si{\tesla}] the magnetic field strength.

To determine $C_\text{mag}$,  we rotate the field by
\SI{90}{\degree} very quickly. Therefore,  we can initially assume the
bacterium to be orthogonal to the magnetic field
$\phi(0)=\nicefrac{\pi}{2}$. Solving the differential equation then
yields
\begin{align}
  \phi(t) &= 2 \cot^{-1}  \exp{\left( \frac{mB}{f} t \right)} \nonumber\\ 
  &\approx \frac{\pi}{2} \exp{\left( -0.85\frac{mB}{f} t \right)} \, .
	\label{eq:diff_eq_solution}
\end{align}
This approximation is better than \SI{0.065}{rad} (see Appendix~\ref{sec:appendix_cotan}). The angle $\phi$
can be  estimated indirectly from the measured scattering as described
by Eq.~(\ref{eq:gdef}) if we assume that the bacteria remain in
the plane of rotation ($\beta=0$). The settling time of this transition period
is characterized by the time constant $\tau = f/mB$. As in our earlier
work~\cite{Pichel2018}, we scale the response time to the magnetic
field and introduce a general rotational velocity parameter $\gamma$ (\si{\radian/Ts})
\begin{equation}
	\gamma = \frac{m}{\pi f} = \frac{1}{\pi \tau B} \, . \nonumber
	\label{eq:gamma}
\end{equation}

From the response time an indication of the
ratio between rotational drag coefficient and magnetic moment of
single cells can be obtained~\cite{Erglis2007,Pichel2018}. The rotational drag
coefficient can be estimated from the bacteria shape using a spheroid
approximation, slender body theory or measurement of macroscopic
models in glycerol~\cite{Steinberger1994, Pichel2018}.

\subsection{Brownian motion}
\label{sec:brownian}
When we remove the magnetic field, magnetotactic bacteria will quickly
reorient in a random orientation distribution via Brownian motion, and
possibly via flagellar motion. For the same reason, the bacteria will
not align perfectly along the magnetic field. The effect of Brownian
motion on the alignment decreases with increasing fields, so we may
expect $C_\text{mag}$ to be field-dependent. Let us first consider the
effect of Brownian motion.

The probability distribution of finding magnetotactic bacteria tilted at an angle of
$\phi_0$ from the magnetic field direction $b(\phi_0)$ is determined
by the ratio of magnetic ($-mB\cos(\phi)$) and thermal energy ($kT$)
according to the Boltzmann distribution~\cite{Rosenblatt1982}. We should take into account that energy
states for a specific value of $\phi$ exist in a full revolution
around the field axis ($\beta=0 \dots 2\pi$). Therefore
\begin{align}
  b(\phi_0) =& \frac{\int_0^{2\pi} e^{a\cos\phi_0} \sin(\phi_0) d\beta  }
               {   \int_0^\pi \int_0^{2\pi}  e^{a\cos\phi}
             d\beta d\phi }
               \nonumber \\
  =& \frac{a}{2 \sinh(a)} \sin(\phi_0) e^{a \cos(\phi_0)} \, , \nonumber
     \label{eq:brownian_rotation}
\end{align}
where $a=mB/kT$, with $k$ (\si{\joule\per\kelvin}) being the Boltzmann
constant and  $T$ (\si{\kelvin}) the temperature.

To achieve a first-order approximation, we assume that the scattering factor is
proportional to the projection of the bacteria shape onto the light
direction. Defining $\alpha$ as the angle between the bacteria long
axis and the light path, the scattering factor in Eq.~(\ref{eq:gdef})  becomes
\begin{align}
  g(\alpha)=1-\lvert\sin(\alpha)\rvert \, . \nonumber
\end{align}
The angle $\alpha$ is the combined result of the angle between the
light and the field direction $\theta$ and the angle between the
bacteria and the field $\phi$.  One can show that the relation between
$\alpha$ and these three angles is
\begin{align}
  \cos(\alpha)=-\sin(\theta)\sin(\phi)\cos(\beta) +
  \cos(\theta)\cos(\phi) \, , \nonumber
\end{align}
resulting in the following expression for the scattering factor:
\begin{align}
  g(\theta,\phi,\beta)=1-\sqrt{1-\cos(\alpha)^2} \, . \nonumber
\end{align}
The average scattering factor can be obtained by double numerical integration,
first over all values of $\beta$ and then over the distribution of
$\phi$
\begin{align}
  <g(\theta)>=\int_0^{\pi} g(\theta,\phi)b(\phi) d\phi \, . \nonumber
\end{align}

\begin{figure}
  \centering
    \includegraphics[width=\widefigurewidth]{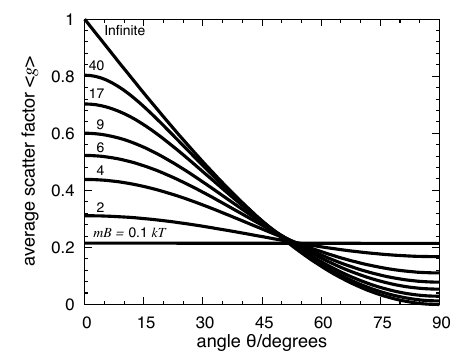}
    \caption{Calculation of the average scattering factor as a function
      of the angle of the field with respect to the light incidence
      for varying values of the product of the magnetic moment of the
      magnetosome chain $m$ and the applied field $B$, expressed in units of
      $kT$ at room temperature. When all bacteria are perfectly
      aligned ($mB/kT=\infty$), the average scattering factor is
      inversely correlated to the projection cross section of the
      bacteria on the light path ($g=1-\sin(\theta)$). At lower
      fields, the loss of alignment reduces the angular dependence,
      which disappears for $mB<kT$.}
    \label{fig:gAvVsTheta}
\end{figure}

The numerical integration was performed in Python, the source code of
which is available as Supplementary Material
(angular.py). Figure~\ref{fig:gAvVsTheta} shows the resulting average
scattering factor as a function of the applied field angle for varying
energy product $mB$.  At an energy $mB$ well above \SI{40}{kT}, the
angular dependence approaches a $1-\sin(\theta)$ relationship.

Assuming a dipole moment of \SI{0.25}{\femto\ampere\square\meter} as
reported in our earlier work~\cite{Pichel2018}, $mB$=\SI{40}{kT}
corresponds to a field of about \SI{0.7}{mT}. Therefore fields on the
order of a few mT may be sufficient to obtain the maximum value of
$C_\text{mag}$.

When the field is removed, the scattering factor is $g_0 =\num{0.2146}$. In
this case, the intensity at the detector is $I_0=g_0I(0)+(1-g_0)I(90)$,
which we can relate to the average $O_\text{D}$ of the suspension
\begin{align}
  O_\text{D} =& \log(\frac{I_\text{ref}}{I_0})    \nonumber \\
  =& -\log\left( g_0 10^{-O_{\text{D}\parallel}} +
  (1-g_0)10^{-O_{\text{D}\bot}} \right) \, .
  \nonumber
\end{align}

In the above, we ignored the disturbing force caused by
the flagella. Flagellar motion is complex~\cite{Pan2009}, making the disturbing force 
difficult to calculate. However, we know that, in natural conditions,
magnetotactic bacteria can use the Earth's magnetic field of about
\SI{50}{\micro T} to navigate. In this low field, $mB$ is only
\SI{3}{kT}. If the stochastic energy provided by
the flagella is much greater than this value, the bacteria would not be
able to follow the field. This suggests that, for fields on the order
of mT, flagellar motion can be ignored.

\section{Method}
\label{sec:design}

\begin{figure*}[t]
  \centering
  \includegraphics[width=\textwidth]{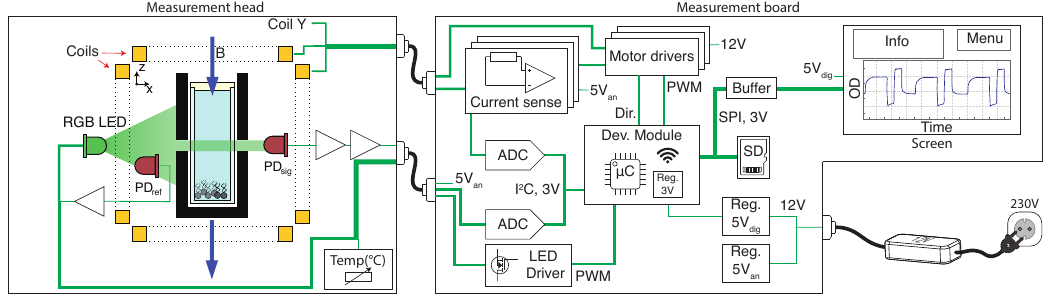}
  \caption{Diagram of the system: Measurement head: LED, refdiode, diode, amplifier stages, coils, cuvette, bacteria. Measurement board: outlet to 12 supply, \SI{12}{V} to \SI{5}{V} analog and digital, microcontroller, AD converter, LED driver, motor shields, current sense, HDMI cable, coil cable, SD card, WiFi. User interface: resistive touchscreen. Measurement board and user interface reside in the same housing.}
  \label{fig:DiagramODmeter}
\end{figure*}


Our MagOD system is an alternative to the modified commercial optical
density meters  currently used in magnetotactic bacteria
research. It should therefore use compatible cuvettes and have
comparable specifications. The preferred wavelength at which absorbance
is measured is approximately \SI{600}{nm}, and the maximum absorbance is
approximately \SI{1.4}{}~\cite{Lefevre2009}. Intensity variations due
to the change in direction of the magnetic field can be as high as
\SI{200}{\percent}, but values as low as \SI{2}{\percent} have been
reported~\cite{Song2014}. Although fields up to \SI{70}{mT} are
applied~\cite{Schuler1995},  there are indications that saturation
occurs as low as \SI{2}{mT}~\cite{Song2014}. Our design therefore should have  a wavelength of approximately \SI{600}{nm}, an
absorbance range of at least \SI{1.5}{}, intensity resolution better
than \SI{1}{\percent}, and a magnetic field above \SI{2}{mT}.


The MagOD system has two main components, see
Figure~\ref{fig:DiagramODmeter}. The cuvette filled with the sample to
be investigated is inserted into the \emph{measurement head}, which
holds the light source and photodetector circuit boards, the three
coil sets and additional sensors (such as temperature). The
measurement head is connected to the \emph{measurement board}, which
holds the analog-to-digital (AD) converters, the drivers for the
magnetic field generation, and the light source.  A microcontroller is
mounted on the measurement board, which is connected over the board to
the AD converters, the data storage card, and a touchscreen.

The design files for the hardware and software components are
available at
\href{https://github.com/LeonAbelmann/MagOD}{github.com/LeonAbelmann/MagOD}.

\subsection{Measurement head}

We designed the measurement head to be as compact as possible to keep the volume and power consumption low. 
The dimensions of the standardized cuvette \num{12.5}$\times$\num{12.5}$\times$\SI{45}{mm}) determine the size
of the coil system, which essentially sets the outer dimensions of the
measurement head. The circuit boards for the light source and sensors are embedded
inside the coil system, with sensors located as close to the cuvette as possible.

\subsubsection{Mechanical}
As the measurement head carries all components, it is a complex
structure that must be modified regularly to accommodate changes in
component dimensions and added functionality. Therefore we decided to
fabricate the structure by 3D printing so that modifications can be
easily implemented. Printing in metal is still prohibitively
expensive, so the measurement head itself cannot act as electromagnetic
shielding. Instead, shields will have to be implemented on the circuit
boards. However, it is possible to 3D-print with black nylon to shield the
photodetector  from external light and to allow the parts to be
easily disinfected with a \SI{70}{\percent} ethanol/water solution.

The measurement head consists of more than a dozen parts. The design
is parameterized using the open-source OpenSCAD language, so that
dimensions can be easily changed. The source files are available at
\href{https://github.com/LeonAbelmann/MagOD}{github}.

\subsubsection{Coil system}
We can choose between permanent magnets or electromagnets with or
without cores to apply a magnetic field.  As the field to be applied
is relatively low, and the field direction needs to be changed rapidly
to monitor the response of the bacteria, electromagnets without cores
provide a simple, light weight solution. An additional advantage is
that the field is directly proportional to the current, and there is
no hysteresis, so no additional magnetic field sensors are
required. The disadvantage of not having a core is that the maximum
field is limited to a few mT. Higher fields can only be applied for
short periods of time, limited by coil heating.

The magnetic field is generated by three orthogonal sets of two coils located on
either side of the sample. The dimensions of the coils are more or
less defined by the cuvette height, but we can choose the wire
diameter to optimize the number of windings $N$. The field in the coil
is proportional to the product of the current $I$ and $N$. The resistance $R$ of the coil scales approximately with $N^2$ for fixed coil dimensions. Therefore, the power dissipated in the coils ($I^2R$) is relatively independent of the number of windings for a given field strength.
The inductance of the coil $L$ scales with $N^2$, so the
cutoff frequency (proportional to $R/L$) is also fairly independent
of the coil wire diameter. The choice of wire diameter
is therefore  determined mainly by the availability of power supplies,
specifications of H-bridges and current ratings on connectors. Table~\ref{tab:coils} shows the specifications of two
commercially available coils (Jantzen Audio 000-1235 and
000-0996) that are both suitable for our application. The number of turns was estimated from the coil resistance
(using literature values for wire resistance) and the coil
inductance~\cite{Lane2001}. Our MagOD system incorporates the coil with the higher number of windings
(\num{996}) to benefit from the substantially lower currents, but at the
expense of a slightly higher cutoff frequency and higher power consumption.

\cmidrulewidth=.03em
\renewcommand{\arraystretch}{1.1} 
\begin{table}
  \caption{Examples of coil specifications.}
  \label{tab:coils}
  \begin{ruledtabular} 
     \begin{tabular}{@{} l c c  l @{} }
       Jantzen Audio coil no. & {1235} & {0996} & \\
       \cmidrule(lr){2-3} 
       Wire gauge & 18 & 22 & AWG \\
       Wire diameter & 1.0 & 0.64 & mm\\
       Resistance & 21 & 53 & \SI{}{m\ohm/m}\\
       \\
       Inner diameter & 42 & 42 & mm\\
       Outer diameter & 57 & 53 & mm\\
       Height & 21 & 21 & mm\\
       Inductance & 0.94 & 2.9 & mH\\
       Resistance & 0.5 & 2.1 & \SI{}{\ohm}\\
       \\
       Cutoff frequency & 85 & 115 & Hz\\
       Windings\footnote{estimated from resistance and inductance} &
                                     \SI{80(3)}{}  & \SI{140(7)}{} \\
       Current for \SI{1}{mT} & 0.9\footnote{estimated from number of
                                windings} &
                                            0.5\footnote{from Figure~\ref{fig:BVsICoil}} & A \\
       Voltage for \SI{1}{mT} &  0.44 & 1.1 & V\\
       Power  & 0.4 & 0.5 & W \\
       \end{tabular}
       \end{ruledtabular}
\end{table}

\subsubsection{Temperature sensor}
Electromagnets---especially those without cores---produce heat as
a byproduct of the magnetic field.  In the absence of active cooling,
the temperature of the sample under investigation can rise quickly.
This is especially problematic when one is  working with
micro-organisms. Therefore, it is important to monitor the temperature
of the cuvette. The best option would be to insert a temperature
sensor into the cuvette, but this method is cumbersome and risks exposing the sample to the outside air. The temperature of
the coils can be estimated from their resistance, but that would
overestimate the temperature of the cuvette. Therefore, we chose to
mount a simple negative temperature coefficient (NTC)  sensor in the housing as close as
possible to the cuvette.

\subsubsection{Light source}
Ideally, the absorption pattern of a specimen is measured over a broad
range of wavelengths. Most optical density meters use a wide-spectrum
xenon flash lamp combined with a monochromator. However, this is not only a
rather power-hungry, bulky solution ($>$\SI{10}{W}, \SI{20}{mm}), it is also
overkill for observing magnetotactic bacteria. Instead, we
chose an RGB LED as source. LEDs are simple to control, can be
mounted close to the cuvette, operated in continuous mode and
 easily adjusted in intensity using pulse width modulation
(PWM). However, the wavelength  cannot be chosen continuously, whereas the
wavelength spectrum is determined by the LED type. Moreover, the
wavelength bandwidth per color is rather large (\SI{25}{nm} compared
to \SI{5}{nm} for monochromators). Finally, the light intensity of a
LED is low compared to that of xenon lights or lasers. Based on the
manufacturer's data, the
LED power  in our current implementation is approximately \SI{0.2}{}, \SI{0.1}{} and \SI{0.7}{\micro
  W/mm^2} for \SI{645}{} (red), \SI{520}{} (green) and \SI{460}{nm}
(blue) light, respectively. This is sufficient for most
suspensions of magnetotactic bacteria.

The LED has a non-diffuse housing such that the light output in the
direction of the sample is optimal. The LEDs can easily be exchanged, for instance
for  a yellow or UV LED, because they are mounted on a separate
board.

The LED is mounted in common anode configuration such that (i)~it can
be driven by NPN MOSFETs and (ii)~the supply difference between the
LED (\SI{5}{\volt}) and the microcontroller (\SI{3.3}{\volt}) is
inconsequential. The frequency of the PWM signal is well above the
cutoff frequency of the photodetector amplifier. Since the brightness
of LEDs decreases with time, we monitor the LED intensity. For this
purpose a photodiode is placed in close vicinity before the light
enters the cuvette.

\subsubsection{Photodiode}
The light passing the cuvette can be detected with photomultiplier
tubes, avalanche photodiodes and silicon
photodiodes~\cite{Hergert2014}. Photomultipliers are highly sensitive,
but are also quite bulky, require high voltages and perform less well
at long wavelengths. Avalanche photodiodes are also very sensitive,
but suffer from nonlinearity, noise, a high temperature dependence,
and require high voltages to operate. As the transmission of light
through most magnetotactic bacteria suspensions is high and we work at
low acquisition frequencies, the sensitivity of silicon photodiodes is
sufficient and allows us to take advantage of its small form factor,
linearity and ease of operation.  We used the more light-sensitive
large-area photodiodes to boost sensitivity. The diode is operated in
photovoltaic mode. In this mode, the bias voltage is zero, so the dark
current, which is highly temperature-sensitive, is minimized.

In our MagOD implementation, the photodiode current is amplified by a
two-stage operational amplifier (opamp) circuit because the signal is too weak for a single-stage amplifier. The first stage is a current-to-voltage converter. A
low-noise JFET opamp is applied because this type of opamp has
a low input current offset, which reduces DC errors and noise at the
output. The first stage has the largest amplification in order to minimize the
amplification of noise. The amplifier circuit is located directly behind
the photodiode inside an electromagnetic protective casing, so that noise picked up by the
cabling to the main board is not amplified and interference is minimized.
 


\subsection{Measurement board}

Placing the photodiode amplifier directly behind the photodiode is an
effective way to suppress interference. We have the option to transport the
amplified photodiode signal directly to the measurement board or to
include the analog-to-digital (AD) converter next to the amplifier in the measurement head
and convert the analog signal into a digital one. The analog option has the
advantage of a small form factor for the circuit board and better
access for testing. The digital option is more robust against
interference and allows simpler cabling.  As the current
implementation of the MagOD system is very much a development
instrument, we chose to move the AD converters to a separate
measurement board, together with the microprocessor and other
peripherals.

\subsubsection{AD converter}
The measurement board has two AD converters to
read out the various analog signals on the system. As the measurements
are normally performed on a longer time scale, we chose converters
that are able to perform measurements with a sampling rate of up to
\SI{860}{\hertz} and have integrated anti-aliasing filters. A 16-bit
resolution provides an upper limit to the absorbance of
4.8, which is more than sufficient. In practice, the absorbance range
is limited by stray light scattering around the sample.

The AD converters have a free-running mode, which performs
measurements at an internally defined clock rate. A data-ready pin
functions as an external interrupt such that the microcrontroller can
be freed for other tasks while waiting for the AD converter to
finalize its acquisition step.

\subsubsection{Microcontroller}
As data acquisition rates are low, the MagOD system can be easily
controlled by a microcontroller ($\mu$C). This allows us to benefit from recent
developments in low-cost, versatile $\mu$C development
platforms. Rather than embedding the $\mu$C directly on the
the electronic board, we chose to include it as a development
board. This way, the system can be easily assembled, debugged and repaired.

The current implementation of the MagOD instrument is built around an
ESP32 development board (Espressif Systems, Shanghai, China). The ESP32 $\mu$C has several characteristics
that make it very suitable for this application: It has a small form
factor, a fast 32-bit dual-core processor operating at
\SI{240}{\mega\hertz}, WiFi and Bluetooth as well as several peripheral
interfaces such as SPI and I$^2$C. This $\mu$C is very popular,
and numerous dedicated libraries, examples and
discussions are available on Internet fora. Additionally, there is a plugin for the
Arduino IDE, and many libraries are natively compatible, so even
inexperienced developers can start with little effort.


\subsubsection{Display}
A resistive touchscreen enables the user to control the system
 with or without protective gloves. Additionally, the screen
provides the user with information on the current and past states of
the measurement and levels of the signals. Line drivers on the main
board ensure that communication is reliable.

\subsubsection{Storage}
The acquired data and recipes are stored on a secure
digital (SD) card. These cards are readily available in a variety of
capacities, are widely applied in DIY projects, and are replaceable in
case of a damaged card. The SD card can be interfaced to the $\mu$C in
the SPI, the 1-bit SD, and the 4-bit SD modes. Although data transfer
is faster in the 4-bit SD mode, we chose the SPI mode because it is
well supported and the write speed is sufficient for our
purpose. However, the write time to an SD card over an SPI interface
using the ESP32 $\mu$C is unpredictable, with SD card-induced peaks in write time of at least \SI{50}{ms}. Fortunately, the
ESP32 has two cores, so unpredictable processes such as access to the SD
card, reaction to touchscreen input and WiFi file transfers can be
moved to a separate core.

\subsubsection{Current drivers}

The current through the coils must be controlled to obtain a specific magnitude of the magnetic field. 
We use PWM and benefit from the fact that the high inductance of the
coil provides a low-frequency, low-pass filter for free. 
The use of PWM minimizes power dissipation in the supply, but results in a current
ripple and consequently a ripple in the magnetic field. This ripple can be
suppressed by choosing a sufficiently high PMW frequency. We use
commercial motor drivers (Cytron MD13S) because they are specialized for driving high
currents through a coil in two directions based on a simple two-wire
control. The currently employed drivers work with frequencies up to
\SI{20}{\kilo\hertz}, suppressing the ripple by a factor of at least 
100. The drivers can be interchanged by alternative motor drivers with
similar capabilities.

The magnetic field is linearly dependent on the current. However, the
current is not linearly dependent on the PWM duty cycle, as the internal
resistance of the coil will vary due to temperature changes. A
precise measurement of the current is necessary to close the loop and
to assess the applied magnetic field. Therefore, a shunt resistor is
placed in series with each coil. The voltage drop over this resistor
is amplified using a current sensing amplifier and digitized with
the AD converter. The measured signal can serve either to determine
the true current or it can be applied in a feedback loop to compensate for coil heating.


\subsubsection{Power supply}
The measurement board electronics operate at low voltages
(\SI{3}{} or \SI{5}{V}). However, the magnetic coil system is
preferably operated at higher voltages to limit the currents and
subsequent requirements for cabling and connectors. For reasonable
winding wire diameters, the currents are in the range of a few ampere
and the resistance of the coils is on the order of a few ohm. Therefore,
we selected  \SI{12}{\volt} for the main on-board supply, for which a
wide range of external power supplies is available and that even allows
for operation from a car battery while in the field.

In our MagOD implementation, the three coil sets have a
combined resistance of \SI{4.2}{\ohm} at room temperature. The maximum current is close to \SI{3}{A} with \SI{12}{V}. This maximum current simultaneously through each coil set  would require a maximum power supply of  \SI{120}{W}.

The analog and digital circuits have a separate \SI{5}{V} supply line to
prevent noise originating from the switching nature of the digital
circuitry from interfering with the measurement. The analog \SI{5}{V} supply is
built using an ultra-low-noise linear regulator, whereas the digital
\SI{5}{V} is built with a switching regulator. The latter is more efficient,
but produces inherently more electronic noise. The \SI{3}{V} needed for the
$\mu$C originates from a linear regulator integrated on the
development board.


\subsubsection{Enclosure}
The device is enclosed in a laser-cut plastic housing.  Plastic was chosen because it does not block the WiFi signal. There is no need to deal with interference signals because the measurement signal
is amplified in the measurement head, and the unshielded sections of
the leads to the AD converter are kept very short.

The design is optimized such that no extra materials are needed for
assembly. Additionally, the parts can be manufactured with a 3D
printer. The source code for the enclosure design is available on
\href{https://github.com/LeonAbelmann/MagOD}{github}.

\subsection{Cabling}
While designing the MagOD system, we envisioned that measurements
could take place inside controlled environments, such as incubators
and refrigerators. Therefore the system was separated into two parts,
connected by cabling. Components that did not have to be on the
measurement head were moved to a separate module. This approach has
the additional complication of requiring cabling and
connectors. To mitigate this drawback, we chose commercially available cabling wherever possible.

For communication with the amplifier boards in the measurement
head, we chose an HDMI cable, which features shielded twisted-pair
wires with a separate non-isolated ground. HDMI cables are ideal for transmitting analog signals with low interference (5~V, signal,
ground). The HDMI interface has evolved through several standards. The
HDMI2.1 + Internet standard has five shielded twisted pairs that can be
used for measurement signals (for instance three photodiodes, NTC and Hall
sensor) and four separate wires that can be used for control signals
(three LEDs).  The connectors on the main board, amplifier boards and
motor drives are standard Molex connectors. The coils are connected to
standard measurement leads with banana connectors. The connection from
the banana plugs to the measurement head is based on a Hirose
RP 6-pole connector, which is the only cable that cannot be purchased
in assembled form.

\subsection{Software}
Most modifications to the MagOD system will be made at the software level,
which will  be performed primarily by students. Generally speaking, (electrical)
engineering students and many hobbyists are skilled in programming 
Arduino development boards. Therefore, the $\mu$C (ESP32) was
programmed in the same way as an Arduino project, using \CC{} and the
native Arduino IDE both as compiler and uploader. This has the major
advantage of posing a negligible entry barrier for inexperienced $\mu$C
programmers.

The disadvantage of the Arduino IDE is that it is not ideal
for larger projects. The current implementation  already exceeds
5000 lines of code. To partially relieve this issue, the code was set up in a highly modular way to assist new programmers in navigation, using only one main source file
(.ino, .h) of 1000 lines, and several local library source files
(src/*.cpp), e.g.\ for screen access, readout of the AD converter, writing to Flash
memory, and WiFi access. The source code can be found on
\href{https://github.com/LeonAbelmann/MagOD}{github}.

The data is collected on the SD card and transferred over WiFi in a
format that can be easily imported and displayed in a spreadsheet
program. For more advanced analysis, Python scripts are available on \href{https://github.com/LeonAbelmann/MagOD}{github}.


\section{Results}
\label{sec:results}

We have analyzed the performance of our MagOD system,
and shall compare it to a commercial spectrophotometer in the first part of
this results section. To illustrate the possibilities of our novel instrument, we
provide three examples in Section~\ref{sec:applications}.


\subsection{Performance}
\label{sec:performance}
Several iterations of MagOD systems have been realized based on the
design considerations described in Section~\ref{sec:design}. We expect that more
iterations will follow, not only by our team but also by others in the
field of magnetotactic bacteria. The most recent implementation can be
found on
\href{http://github.com/LeonAbelmann/MagOD}{github}. We measured the
performance of our  MagOD meter (version
2) with respect to its optical and magnetic components to provide a
baseline for future improvement.

\subsubsection{LED and photodetector}

\begin{figure}
  \centering
    \includegraphics[width=\widefigurewidth]{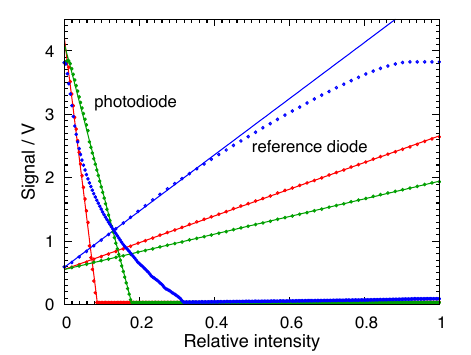}
    \caption{Photodiode and reference diode signal versus LED
      duty cycle (proportional to effective power). Note that the
      signal on the photodiode decreases with increasing light
      intensity due to the particular two-stage design of the
      amplifier. The reference diode amplifier is based on a
      conventional one-stage design. The photodiode signal is linear
      with the duty cycle for the red (\SI{645}{nm}) and green
      (\SI{520}{nm}) LEDs, but not for the blue (\SI{460}{nm})
      LED. Similarly, the reference diode signal is linear with the
      duty cycle of the red and green LEDs, but not for the blue. The
       parameters for the linear fits are listed in
      Table~\ref{tab:VdioVsLPLEDfits}.}
    \label{fig:VdiodeVsPLED}
\end{figure}

\paragraph{Photodetector sensitivity.}
Our MagOD system is equipped with a three-color LED, which allows
selection of three wavelengths (peak intensities at \SI{645}{},
\SI{520}{} and \SI{460}{nm}), either individually or in combination. The
LEDs are individually driven by a PWM voltage to adjust their
intensity, for instance to match the transmission of light through the liquid in the
cuvette. A reference photodiode is mounted adjacent to the LEDs, which
captures a small fraction of the LEDs' light to monitor variations in
the emitted light intensity. Figure~\ref{fig:VdiodeVsPLED} shows the
signal of the detector and reference photodiodes as a function of the
average LED power for the three  wavelengths. The light
pattern is shown in Figure~\ref{fig:LEDProjection} in the appendix,
and a video is provided in the Supplementary Material (MagODLEDProjection.mov).

Space restrictions compelled us to design the two-stage amplifier such
that the output decreases with increasing LED power. The reference
photodiode, which has only one amplifier stage, has an increasing
output with increasing intensity.

The relation between output voltage and intensity is linear for the
red and green LEDs, but not for the blue LED at higher
intensities. Measurements with liquids of different absorbance confirm that the sensitivity
to blue light drops at high intensities of  incident
light, see Figure~\ref{fig:SignalVsPower} in the appendix. Therefore, the blue LED should   be used only for accurate
absorbance at low incident power, i.e.\ for signals above \SI{2}{V}. At low
intensity, the sensitivities of the red and blue channels are
approximately equal, and twice as high as that of the green channel for the
chosen combination of LED and photodetector. However, the sensitivity of the
reference photodiode to red and blue light is clearly
different. This again may be related to the placement of the diodes in
the LED housing.

The linear fits to the data are listed in
Table~\ref{tab:VdioVsLPLEDfits}. The offsets are in agreement with the
manufacturer's specification of the ADS115 of \SI{4.096}{V}. 

\cmidrulewidth=.03em
\renewcommand{\arraystretch}{1.5} 
\begin{table}
  \caption{Linear fits to measurements shown in
    Figure~\ref{fig:VdiodeVsPLED}. The blue LED has a nonlinear
    response and is not tabulated.}
  \label{tab:VdioVsLPLEDfits}
  \begin{ruledtabular}
    \begin{tabular}{@{}lcccccc@{}}
      &  & \multicolumn{2}{c}{photodiode}   & \multicolumn{2}{c}{reference diode} \\
      \cmidrule(lr){3-4}  \cmidrule(l){5-6}
      LED & $I_\text{max}$ & offset & slope & offset & slope \\
      & {mA} & V & V/$I_\text{max}$ & V & V/$I_\text{max}$ \\
      \hline
      Red & \SI{20(2)}{} & \SI{3.159(5)}{} & \SI{-15.28(4)}{} &
                                                                \SI{0.545(3)}{} & \SI{3.305(5)}{}\\
      Green & \SI{20(2)}{} & \SI{3.168(2)}{} & \SI{-6.696(9)}{} &
                                                                  \SI{0.516(1)}{} & \SI{1.902(2)}{}\\
      Blue &  \SI{20(2)}{} & & & \SI{0.532(1)}{} & \SI{6.152(5)}{}\\
    \end{tabular}
  \end{ruledtabular}
\end{table}

\begin{figure}
  \centering
    \includegraphics[width=\widefigurewidth]{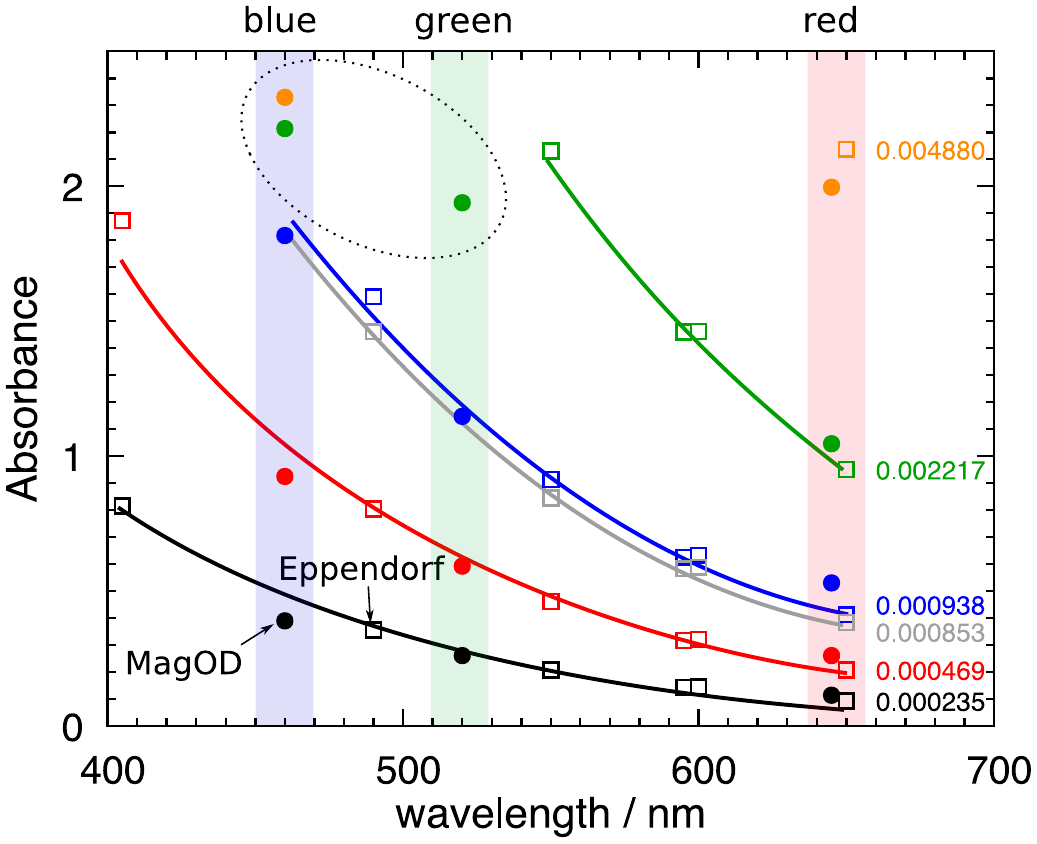}
    \caption{Absorbance relative to water measured with our MagOD
      meter (solid circles in the colored bands) for the three LEDs
      compared to the optical density measured by an Eppendorf BioPhotoMeter
      (open squares) measured as a function of  wavelength. We used a
      range of dilutions of a water-based ferrofluid  (FerroTec EMG 304), the dilution factor of which is indicated on the
      right. The data points for every dilution are indicated by a
      line to guide the eye. The difference between our MagOD meter
      and the commercial instrument is greater than the estimated
      measurement error, but less than \num{0.2} for absorbances below
      \num{2}. Above this value, the estimate is unreliable (data points
      inside dotted loop). The absorbance of the blue LED is
      systematically lower than that of the commercial instrument. The maximum
      absorbance measured was \num{1.82}, which is slightly lower than for
      the commercial instrument (\num{2.14}). The measurement uncertainty is less than the symbol size and is therefore omitted for
      clarity.}
    \label{fig:EppendorfComparison}
\end{figure}

\paragraph{Absorbance validation.}
To validate performance with respect to standard photospectroscopy
measurements, we compared our MagOD system with a commercial optical
density meter (Eppendorf BioPhotoMeter
Plus). Figure~\ref{fig:EppendorfComparison} shows the absorbance ($O_\text{D}$) 
relative to water as a function of the wavelength of the light for a
range of dilutions of a suspension of magnetic nanoparticles (FerroTec
EMG 304). The transmission of light measured by our MagOD meter was
averaged for a range of photodiode intensities ranging from zero
to saturation. For the blue LED, however, care was taken to measure
only at low intensities, where the response is linear, see
Figure~\ref{fig:VdiodeVsPLED}.

As expected, the absorbance increases with increasing nanoparticle
concentration as indicated on the right-hand side of the graphs. The absorbance
increases with decreasing wavelength, which is in agreement with the
observation that the solution has a brown appearance. Care was taken
to determine the accuracy of the measurement as accurately as
possible. At this precision level, it is clear that our novel MagOD meter
and the commercial instrument deviate, albeit never
more than \num{0.2} for absorbances below \num{2}. Above this value,
the deviation becomes considerable, see data points inside dotted loop in Figure~\ref{fig:EppendorfComparison},
probably due to light scattering onto the photodetector through
other paths.

The blue LED seems to systematically underestimate the absorbance,
which may be related to the fact that the response of the detector is
ill-defined. The maximum absorbance is comparable to that of the commercial
instrument. We therefore conclude that our MagOD instrument is
satisfactory as a conventional absorbance meter, especially its red
and green channels.

\begin{figure}
  \begin{center}
    \includegraphics[width=\widefigurewidth]{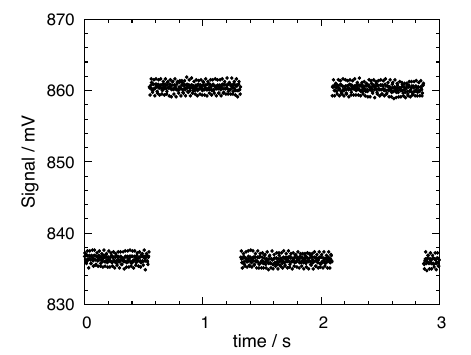}
    \caption{Detector photodiode voltage sampled by the AD
      converter at a rate of \SI{860}{samples/s} while the red LED
      power is modulated by \SI{0.4}{\percent}. The effective
      sample rate was \SI{323}{samples/s}. No
      transitions between the levels can be observed, so the time
      response of the detector photodiode is better than
      \SI{3.1}{ms}.}
    \label{fig:ModulatedSignal860}
  \end{center}
\end{figure}

\paragraph{Time response and noise level.}
The ADS1115 AD converter has a maximum sampling rate of
\SI{860}{samples/s}, which means a sampling time of \SI{1.2}{ms}. Figure~\ref{fig:ModulatedSignal860} shows a
time sequence of the sampled photodiode signal at that rate. The red
LED was switched on and modulated from \num{46} to \num{47} bits on a
full range of \num{255} (relative intensity approximately \num{0.18})
every \num{250} samples. The total acquisition of \num{1300} samples
took \SI{4023}{ms}, so the effective sample rate was only
\SI{323}{samples/s}. The reduction in data rate is due to
communication overhead with the AD converter, and could be
optimized.

The data in Figure~\ref{fig:ModulatedSignal860} shows two clear
levels, with no measurement points in transition from one to the
other. Therefore, we can safely conclude that the response of our MagOD
meter at the highest sample rate is better than \SI{3.1}{ms}. This is
in agreement with the filter applied in the feedback loop of the
amplifier, which has a \SI{-3}{dB} point at \SI{800}{Hz}
(\SI{1.25}{ms}).

\begin{figure}
  \centering
    \includegraphics[width=\widefigurewidth]{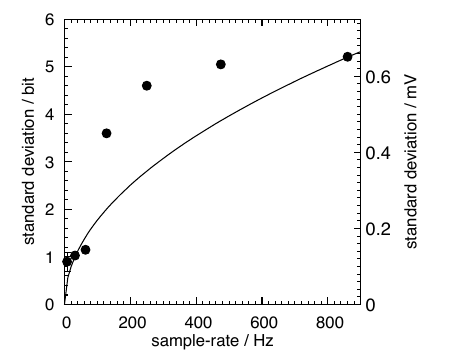}
    \caption{Standard deviation (RMS noise) over \num{1000} samples
      taken by the AD converter of the detector photodiode signal as a
      function of  sample rate. Noise increases with increasing sample rate, but not
      proportional to the square root of the bandwidth (solid
      line). There is a strong increase in
      noise above \num{64} samples/s.}
    \label{fig:noise}
\end{figure}

The ADS1115 has an internal filter that matches the bandwidth, which
can be selected from discrete values of \num{8}, \num{32}, \num{64},
\num{128}, \num{250}, \num{475} and \SI{860}{samples/s}. Therefore, the noise should decrease at lower sample rates.
Figure~\ref{fig:noise} shows the standard deviation of
\num{1000} samples, which is equal to the root-mean-square (RMS) noise as a function of
 sample rate. As expected, noise increases with increasing
sample rate, but much more steeply than can be expected from a white noise
spectrum, i.e.\ noise proportional to the square root of the
bandwidth. There is a strong jump in noise above
\SI{64}{samples/s}, most likely caused by the presence of a
\SI{50}{Hz} cross-talk signal. At \SI{64}{samples/s} and below, the
noise is on the order of \SI{1}{bit} or \SI{125}{\micro V}. The
full range of the detector circuit is \SI{3.1}{V}, which corresponds to
a dynamic range of \SI{88}{dB} or a theoretical upper limit to the
detectable absorbance of \SI{4.4}{}. This compares very favorably to
the commercial Eppendorf system, which has an optical-density resolution  of
\SI{1e-3}{} on full range of approximately \SI{2}{}. Assuming that the
noise level of the Eppendorf system is comparable to the resolution,
this would correspond to a dynamic range of only \SI{53}{dB}.

At \SI{64}{samples/s}, the noise level is \SI{16}{\micro
  V/\sqrt{Hz}}. SPICE\footnote{Simulation program with integrated circuit emphasis} simulations indicate that the theoretical noise level of the
amplifier is on the order of \SI{0.5}{\micro V/\sqrt{Hz}}, indicating that we have
not yet reached the full potential of the electronics. 

\subsubsection{Magnetic field system}

\begin{figure}
  \centering
    \includegraphics[width=\widefigurewidth]{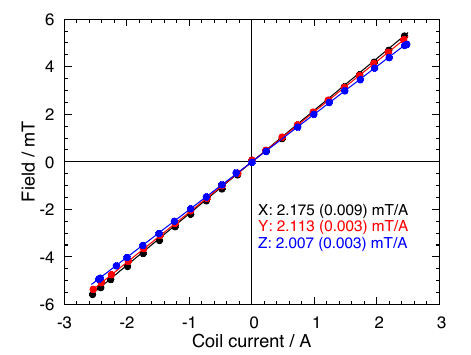}
    \caption{Magnetic field in the center of the cuvette holder as a
      function of average coil current. Fields in $x$, $y$ and $z$
      direction with a field-to-current ratio of
      \SI{2.175(9)}{}, \SI{2.113(3)}{} and \SI{2.007(3)}{mT/A},
      respectively. }
    \label{fig:BVsICoil}
\end{figure}

Figure~\ref{fig:BVsICoil} shows the magnetic field in the center of
the system as a function of the current through each of the three
coil sets. The coils generate approximately \SI{2}{mT/A}, with around
\SI{5}{\percent} variation between the coils. The maximum field that
can be generated is slightly higher than \SI{5}{mT} at full current of
approximately \SI{2.5}{A}. The pulse width of the modulation of
the driver circuits can be set with a maximum resolution of \SI{16}{bit},
corresponding to a theoretical field resolution of about
\SI{70}{nT}. In practice, we operate the PWM at \SI{8}{bit} resolution,
which yields a set-point resolution of about \SI{20}{\micro T}.

As we drive the coils with a PWM signal, the
current through the coils is not constant but follows the modulation
frequency. At zero and maximum current, the ripple is absent. The
ripple has a maximum at \SI{50}{\percent} duty cycle. The filtering
action of the coil system dampens the modulation. At a PWM drive
frequency of \SI{20}{kHz} and \SI{50}{\percent} modulation, we measured
a triangular current signal with a peak-to-peak amplitude of \SI{24(2)}{mA}
on a mean current of \SI{1.2}{A}. Simulations considering only the low-resolution
nature of the coils, with a corner frequency of \SI{115}{Hz}, yield a
theoretical amplitude of \SI{18}{mA}, so there is probably  some
additional capacitive coupling. The current variation corresponds to a
maximum field variation in the field of approximately \SI{50}{\micro T}
or \SI{1.2}{\percent}.

\begin{figure}
  \centering
    \includegraphics[width=\widefigurewidth]{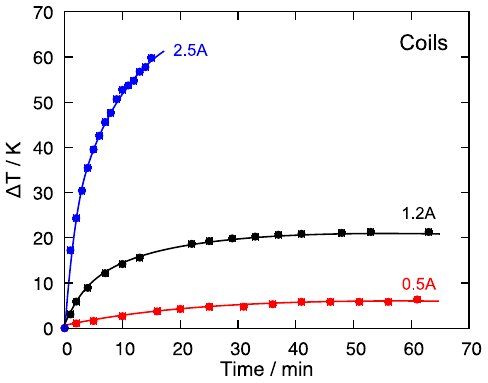}
        \includegraphics[width=\widefigurewidth]{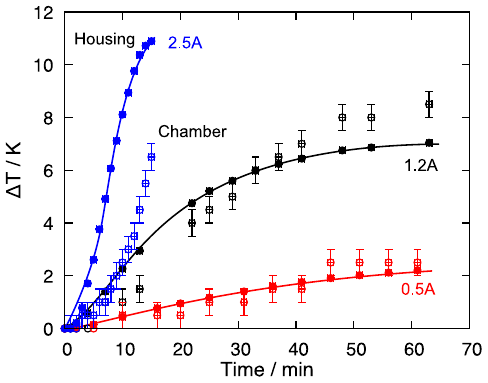}
        \caption{Top: Temperature of the coil system as a function of
          time for different drive currents. At the maximum drive
          current of \SI{2.5}{A}, corresponding to \SI{5}{mT}, the
          coils heat rapidly, and operation should be limited to a few
          minutes. Bottom: MagOD frame temperature (solid symbols) and
          air temperature in the chamber (open symbols) as a function
          of time. At a drive current below \SI{1.2}{A} (\SI{2.4}{mT}), the
          temperature increase of the frame is limited to about \SI{6}{K}.}
    \label{fig:TVsTime}
\end{figure}

At the maximum current of
\SI{2.5}{A}, the coils dissipate about \SI{13.1}{W} each.
As the coil system has no active cooling, the
heating of the sample area can be considerable for prolonged measurement
times. An NTC temperature sensor is mounted
on the body of the measurement chamber to monitor the rise in temperature, see
 Figure~\ref{fig:TVsTime}. We
also measured the temperature in the chamber with a simple alcohol
thermometer for comparison. The temperature of the coils can be estimated from the
increase in coil resistance, assuming the temperature coefficient of
copper (\SI{0.393}{\percent \per K}).

At a drive current of \SI{0.5}{A} (field strength of \SI{1}{mT}), the
heating of the chamber is barely noticeable (about \SI{1}{K/h}). The
average temperature of the coils increases at approximately
\SI{8}{K/h}. With a drive current of \SI{1.2}{A}, the temperature of the
coils increases by \SI{21}{K}. The temperature increase of the chamber 
is substantial, with an initial increase of approximately
\SI{0.25}{K/min} and a subsequent flattening at \SI{7}{} to \SI{8}{K} after \SI{40}{min}.

\subsection{Applications}
\label{sec:applications}
We present four experiments to illustrate the application of our MagOD
meter to analyze magnetotactic bacteria. We measure (1) the
scattering of \emph{Magnetospirillum gryphiswaldense} (MSR-1) bacteria as a
function of their angle to the incident light, (2) their rotational
velocity as a result of a rotation of the external magnetic field on
time scales of seconds and (3) the development of a culture over a period
of several days. The final experiment  measures (4) the velocity
distribution of the unipolar \emph{Magnetococcus marinus} (MC-1) as a
function of time.

\subsubsection{Transmission as a function of angle (MSR-1)}
Using the coil system of our MagOD meter, we can apply a field in any
direction in three-dimensional space. This allows us to study the
transmission of light as a function of the orientation of the
bacteria and check the model presented in Section~\ref{sec:theory}.

\begin{figure}
  \centering
    \includegraphics[width=\widefigurewidth]{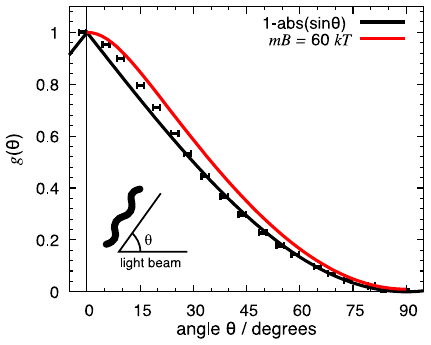}
    \caption{Scattering of a culture of magnetotactic bacteria is dependent on the
      orientation of the external magnetic field; it is highest when
      the field, and hence the bacteria, are aligned parallel to the
      light beam and lowest when the field is aligned orthogonally. By
      normalizing from $[0,1]$, we obtain the angle-dependent scattering
      factor $g(\theta)$, which can be relatively well approximated by a
      sine function. The flattening of the curve around \SI{0}{\degree} can be
      partly explained by additional Brownian motion of the bacteria
      (red curve).}
    \label{fig:gopt}
\end{figure}

For this purpose, a cuvette of MSR-1 bacteria, grown as described in reference~\cite{Pichel2018}, with an optical density of approximately \SI{0.1}{}
was inserted into the MagOD system. We measured the intensity on the
photodetector as a function of the angle of the magnetic field with
steps of approximately \SI{5}{\degree}. The magnetic field varied with
angle, but was always greater than \SI{1}{\milli\tesla}.  As the optical
density of the sample fluctuated continuously due to activity and
sedimentation within the cuvette, we performed the measurement  \num{20} times. The resulting curves were normalized to a range of \num{0} to
\num{1} and averaged to obtain the  angle-dependent scattering factor
$g(\theta)$ shown in Figure~\ref{fig:gopt}.

The simple inverted sine model discussed in Section~\ref{sec:theory} fits
surprisingly well. The strongest deviation is around the parallel
alignment, which is not surprising. The MSR-1 are not infinitely thin
rods, but spirals. Therefore, the projected area will be less
sensitive to variations in the angle around the long
axis. Additionally, the culture of MSR will have a distribution in
angles (due to Brownian motion and/or flagellar motion), which will round off
the sharp corner at $\theta=0$. The red curve illustrates this effect for
$mB$=\SI{60}{kT}, which still does not fit
the measurement very well. It  therefore seems likely that the actual
bacteria shape, and might also be their distribution, should be included
in the model.

\subsubsection{Response as a function of field strength and time (MSR-1)}
\begin{figure}
  \centering
    \includegraphics[width=\widefigurewidth]{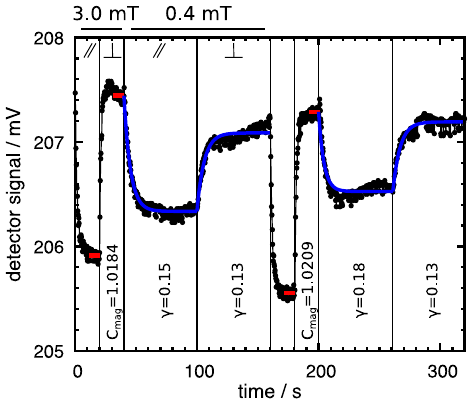}
    \caption{Two cycles of a measurement sequence. The magnetic field
      is alternatively aligned parallel and perpendicular to the light
      beam. For high fields (\SI{3.0}{mT}), we can determine
      $\Delta_\text{OD}$ from the difference between the averages of
      the detector signals (red lines), from which we can calculate
      $C_\text{mag}$. The signal of the growth medium without bacteria was
      \SI{301(1)}{mV}. The difference between the two directions of
      the field drops considerably at low field (\SI{0.4}{mT}),
      whereas the response time increases. These low fields are
      suitable for estimating the time constant $\tau$ from a
      fit to an exponential (blue line). Using the field magnitude, we
      can calculate $\gamma$ (\SI{}{rad/mT.s}). }
    \label{fig:bac_response}
\end{figure}

We most commonly perform measurements in which a sample
of MSR-1 is subjected to field switching between parallel and
perpendicular alignment to the light path and varying field
strengths. Figure~\ref{fig:bac_response} shows the measured response
for a set of field cycles.\footnote{In this measurement, the absorbance
  is high (transmission of light is low) when the field is aligned
  along the light path. This measurement was performed with an older,
  single-stage photodiode amplifier, unlike the measurement shown in
  Figure~\ref{fig:VdiodeVsPLED} taken with the new amplifier, which has
  an inverted response.} At a high field value of \SI{3}{mT}, the field
is switched from parallel to perpendicular alignment after \SI{10}{s}. For the
lower field value of \SI{0.4}{mT}, we can allow  longer reversal times
because coil heating is negligible.

From the difference in detector signals, we can calculate
$\Delta_\text{OD}$ using Eq.~(\ref{eq:deltaOD}).  The signal of the
growth medium without bacteria ($I_\text{ref}$) was \SI{301(1)}{mV}. We can therefore
 calculate $C_\text{mag}$ using Eq.~(\ref{eq:Cmag}).

\begin{figure}
  \centering
    \includegraphics[width=\widefigurewidth]{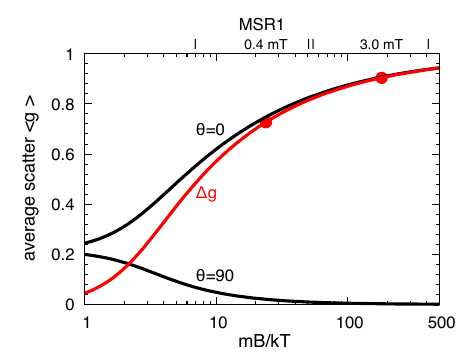}
    \caption{Calculation of the average scattering factor as a function
      of the product of the magnetic moment of the magnetosome chain
      $m$ and applied magnetic field $B$ for the magnetic field
      aligned parallel ($\theta$=\SI{0}{\degree}), and perpendicular
      ($\theta$=\SI{90}{\degree}) to the light path, see
      Figure~\ref{fig:gAvVsTheta}. The difference in the average scattering
      factor between these orientations is indicated by a red line
      ($\Delta g$). From earlier work~\cite{Pichel2018}, we estimated
      that the mean magnetic moment of the magnetosome chain is
      \SI{0.25}{fAm^2}, with a $10-\SI{90}{\percent}$ cutoff of
      the distribution of \SI{0.07}{} and \SI{0.57}{fAm^2},
      respectively. The resulting ranges in $mB/kT$ are indicated on
      the top axis and by red circles for the mean values of the
      moments for the fields used in the experiment illustrated in
      Figure~\ref{fig:bac_response}.}
    \label{fig:gParPer}
\end{figure}

 The difference in transmission between in-plane and perpendicular
 alignment is higher at \SI{3}{mT} compared to \SI{0.4}{mT}. This is
 in agreement with the predicted increase of scattering factor with
 increasing field, see Figure~\ref{fig:gAvVsTheta}.
 Figure~\ref{fig:gParPer} shows the calculated difference in
 scattering factor as a function of the magnetic field scaled to
 $kT/m$. From a previous analysis of MSR-1~\cite{Pichel2018}, we
 estimated that the mean magnetic moment $m$ of the magnetosome chain
 is \SI{0.25}{fAm^2}, with a $10-\SI{90}{\percent}$ cutoff of the
 distribution of \SI{0.07}{} and \SI{0.57}{fAm^2}, respectively. We
 can convert these ranges of moments into the energy ratio $mB/kT$ for
 the two difference field values. Lines on the top axis of the graph
 indicate the ranges, and red circles on the red line indicate the
 mean values. The predicted reduction between the average scattering
 factors (\SI{0.20}{}) at the two field values is less than that
 observed in Figure~\ref{fig:bac_response} (\SI{0.5}{}). The
 discrepancy could originate from the fact that this simple model
 ignores disturbance caused by flagellar motion. Another plausible
 cause is a smaller average magnetic moment $m$ of the magnetosomes in
 this particular sample. Microscopy observations of trajectories of
 other wild-type MSR-1 show alignment for fields of
 \SI{0.4}{mT}~\cite{Pfeiffer2020}, which is in agreement with the
 model.

 In addition to a decrease in step height, the time response also
 decreases with decreasing field. The time constant is estimated from
 a fit of Eq.~(\ref{eq:diff_eq_solution}) to the data using the sum of
 squared errors criterion. The time constant of the transitions to
 \SI{3}{mT} is \SI{1.7(5)}{s}, which is approximately 13 times higher
 than the time constant of \SI{5.4(8)}{s} of the transition to
 \SI{0.4}{mT}. The ratio is rather high, but still within measurement
 error equal to the ratio of the fields, as predicted by the model
 discussed in Section~\ref{sec:theory}.

\subsubsection{Comparison of $C_\text{mag}$ for different instruments}

A major advantage of our novel MagOD system is that it allows us to
standardize $C_\text{mag}$ measurements obtained at different laboratories. It is
not trivial to calibrate instruments by sending around bacteria
cultures because such cultures develop with time and are not stable.
Therefore, a standard instrument is the better
solution. Table~\ref{tab:CmagComparison}  compares 
$C_\text{mag}$ with existing instruments in three different
laboratories. Researchers at CNRS used a modified Cary UV  at \SI{600}{nm}
wavelength, those at Aston University used a modified Thermo Evolution at
\SI{565}{nm} wavelength, and those at Bayreuth University used a Ultrospec 2100 Pro at
\SI{565}{nm} wavelength as well.  The cultures were not the same, so the
results obtained at the above laboratories cannot be compared. Nevertheless, it is very clear
that the difference between the results obtained from commercial instruments and those obtained from our standardized
MagOD vary significantly from laboratory to laboratory.

\begin{table}
  \caption{$C_\text{mag}$ of MSR-1. Comparison of MagOD with commercial
    instruments used at different laboratories (Cary UV: CNRS, Thermo
    Evolution: Aston, Ultrospec 2100:
    Bayreuth). The difference between commercial instruments and our
    MagOD system varies from laboratory to laboratory, which underscores the need
    for a standardized measurement instrument. }
  \label{tab:CmagComparison}
  \begin{ruledtabular}
    \begin{tabular}{@{}l*{7}{c}@{}}
      &  \multicolumn{3}{c}{Cary}   &
          \multicolumn{1}{c}{Thermo} &
           \multicolumn{3}{c}{Ultrospec 2100} \\
      \cmidrule(lr){2-4} \cmidrule(lr){5-5} \cmidrule(r){6-8}
      Commercial Inst. & 2.8 & 3.1 & 2.8 & 2.4 & 2.6 & 2.4 & 2.2 \\
      MagOD Green & 1.9 & 2.1 & 1.9 & 2.3 & 1.5 & 2.3 & 1.5 \\
      MagOD Red    &       &       &       & 2.5 & 1.7 & 1.6 & 1.7 \\
    \end{tabular}
  \end{ruledtabular}
\end{table}


\subsubsection{Long-term growth monitoring (MSR-1)}
When cultivating magnetotactic bacteria such as MSR-1, it is important
to check regularly whether the bacteria remain magnetic. When bacteria are grown under laboratory conditions, random mutation may lead to a culture of
magnetotactic bacteria that has lost the ability to form
magnetosomes~\cite{Popp2014}.  In our lab, MSR-1 are grown in
\SI{2}{mL} tubes. The tubes are closed, and a small headspace of air serves to
ensure a proper reduction of oxygen concentration as the culture
grows.  Even though this method is simple, its major disadvantage is
that we have no information whether the magnetosome formation occurs as we
expect it to. We cannot open the tubes to take samples, because that would let
oxygen in. The better option would be to grow the bacteria in bioreactors that
allow sampling without disturbing the oxygen concentration. Unfortunately, bioreactors are complex,  costly, and provide quantities that far exceed what is needed
for lab-on-chip experiments.

Our MagOD system offers a solution for monitoring the growth of MSR-1
bacteria and the magnetosome by keeping cultures in cuvettes inside
the MagOD meter for long periods. Throughout the growth period, we
continuously measure the absorbance during changes of the external
magnetic field. In this way, we obtain information about the total number
of bacteria as well as their magnetic response.

\begin{figure}
  \centering
    \includegraphics[width=\figurewidth]{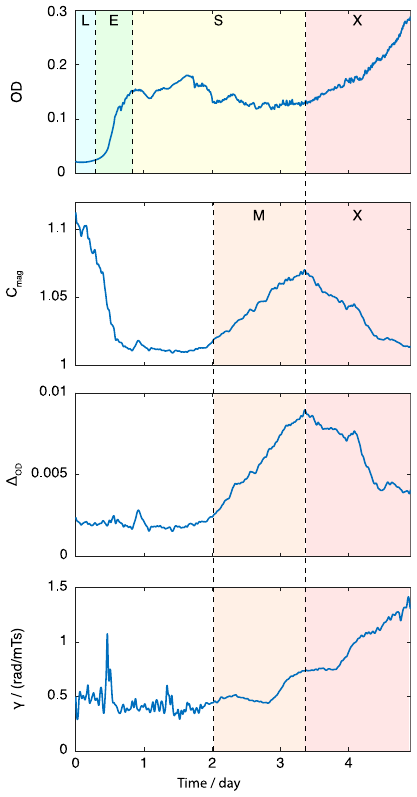}
    \caption{Four bacteria suspension parameters measured over a span of
      five days:  (i) optical density (OD),  (ii) $C_{\mathrm{mag}}$, which
      quantifies the ratio of magnetic to nonmagnetic bacteria, (iii) 
      $\Delta_\text{OD}$, which quantifies the amount of magnetic bacteria,
      and (iv) $\gamma$, which quantifies how strongly the bacteria respond to
      magnetic fields. The following phases can be identified: lag
      L, exponential  E, stationary  S,   magnetic
      growth M, and an undefined phase X.}
    \label{fig:OD}
\end{figure}

We prepared MSR-1 cultures in the conventional manner~\cite{Pichel2018} but, instead
of tubes, we used quartz cuvettes with a PTFE stopper (Hellma QS
110-10-40) to avoid oxygen leakage into the cuvette.  For the
long-term observations shown in Figure~\ref{fig:OD}, the magnetic field was
set to loop through cycles of \SI{100}{\second} consisting of a
vertical field of \SI{1.0}{\milli\tesla} (\SI{20}{\second}), a
horizontal field of \SI{2.9}{\milli\tesla} (\SI{20}{\second}), and a
vertical field of \SI{0.1}{\milli\tesla} (\SI{60}{\second}).

The first transition is at a relatively strong field, thus
guaranteeing reliable estimations of $C_{\mathrm{mag}}$. The second
transition guarantees a relatively high time constant, which is helpful for
 estimating $\tau$ accurately.

Figure~\ref{fig:OD} shows the measured parameters of a sample of magnetotactic bacteria
over a period of five days.  
The optical density, relative ($C_\text{mag}$) and absolute
($\Delta_\text{OD}$) magnetic response and relative rotation velocity
($\gamma$, proportional to the ratio between magnetic moment and
rotational friction coefficient) are plotted from top to bottom.
The optical density is typical for a bacteria growth
sequence. After a lag phase L, a transition into the exponential
growth phase E occurs, followed by the stationary phase S, where the
bacteria concentration remains more or less constant. After three days,
however, the density increases unexpectedly as illustrated in phase
X. Since $\Delta_\text{OD}$ is decreasing, it seems unlikely this
increase is caused by accelerated growth of bacteria or an increase
in scattering due to an increase in intracellular storage granules. 

During the exponential growth phase, $C_\text{mag}$ decreases within
half a day. As $\Delta_\text{OD}$ remains constant, we hypothesize
that the increase in cell density is entirely due to bacteria without
magnetosomes. Only after two days do we see a gradual increase in
magnetic signal due to an increase in the proportion of bacteria with
magnetosomes. With the increase in magnetic signal, $\gamma$ also
increases, hence the magnetic moment of the magnetosome increases
compared to the average bacteria length. The observation for the first
three days would be consistent with the mechanism that, after seeding
with magnetic bacteria, growth first proceeds by an increase in
non-magnetic bacteria. When that growth stops, the bacteria start to
form magnetosomes. This mechanism contradicts electron microscope
observations by Staniland and Yang that magnetosome crystals are
distributed equally over both parts of the divided
cell~\cite{Staniland2010, Yang2001b}. Another hypothesis is that after
cell division, the two daughter cells are shorter and therefore
optically less anisotropic, leading to a reduction in
$C_\text{mag}$. The division must be such that the the absolute change
in absorbance by the two new cells, $\Delta_\text{OD}$, remains
constant. This is far from obvious.

A sharp transition in phase X occurs after approximately \num{3.3}
days. As the density of the culture increases again, the magnetic
response decreases but $\gamma$ continues to increase. This is a
feature we often observe in these measurements(~\cite{Pichel2018a},
appendix B). In this particular case the noise in $\gamma$ decreases,
which was not observed in other experiments. Variations in culture
growth over time between experiments are not uncommon, even under
controlled conditions~\cite{Fernandez2018}. However, we sometimes
observe a cloudiness in the suspension, which may be caused by
aerotaxis or contamination. As we do not shake the suspensions before
measuring like in a standard optical density meter, these clouds may
float in front of the detector and complicate the analysis. It is
possible that, rather than rotating individual bacteria, we rotate the
entire cloud. Clearly, this type of experiment needs to be developed
further.

\subsubsection{Marathon test: MC-1 velocity measurement}

In contrast to MSR-1 bacteria, which reverse frequently, magnetotactic
bacteria of type MC-1 tend to swim for long periods in the same
direction~\cite{Lefevre2009}. This allows us to collect a large number
of bacteria at the bottom of a cuvette simply by applying a vertical
field. After reversing the direction of the field, all bacteria swim
upward in a band-shaped cloud. In the MagOD system, the passing of this cloud
translates to a drop in the photodetector signal. The time between
reversal of the field and the response on the photodetector is a
measure for the velocity of the bacteria. We call this method the
``marathon'' test.

To obtain sufficient bacteria for this experiment, we cultivated MC-1
bacteria in a high-viscosity, agarose-based medium in an oxygen
gradient, as described by Bazylinski~\cite{Bazylinski2013}, but use
low-melt agarose instead of bacto agar. The bacteria form a band in
the reaction tube a few millimeters below the surface of the medium~\cite{Lefevre2016}. The easiest way to free the MC-1 from the medium is to
pipette a small amount from the band and insert this into a cuvette
filled with a low-viscosity growth medium from which the agarose has
been omitted. The transfer of some agarose cannot be avoided,
especially if large quantities of bacteria are desired.

Alternatively, one can place a droplet of agarose with bacteria on one
side and a droplet of growth medium with agarose next to it
so they merge. One can then use a magnet to draw the MC-1  out of the agarose
and into a clean droplet. The disadvantage of this method is that only a limited amount of bacteria  can be collected and that it is difficult to avoid admitting oxygen into the sample.

The method we prefer is to pass the mixture of bacteria
and agarose through a Pasteur pipette filled with a small plug of
cotton. Our assumption is that the cotton breaks up the agarose matrix
and perhaps even captures it. By using compressed nitrogen to push
the growth medium with bacteria through the plug, exposure to oxygen
can be avoided. To  reduce oxygen exposure even further, we performed this
procedure in a nitrogen atmosphere. For this purpose, we simply use a glass
beaker with a paraffin cover through which the Pasteur pipette is
inserted into the cuvette.

The MagOD system is equipped with a 3D magnetic coil configuration, which makes it simple to
apply a vertical field along the cuvette. A field
of \SI{1}{mT} is applied in  positive $z$-direction for \SI{220}{s}
to collect south-seeking bacteria at the bottom of the cuvette. Then
the field is reversed so that the collected bacteria swim upward
towards the photodetector.  We allow the bacteria to swim upward for
\SI{200}{s}, after which the sequence is repeated. The asymmetry in time
ensures that a sufficient amount of bacteria can reassemble at the
bottom of the cuvette. The cloud of bacteria that leaves the
bottom of the cuvette disperses due to a distribution of bacteria
velocities. To keep the peak sharp and intensity variation high, we
reduce the distance between the bottom of the cuvette and the light
path to \SI{2.5}{mm} by using a special insert.

\begin{figure}
  \centering
    \includegraphics[width=\widefigurewidth]{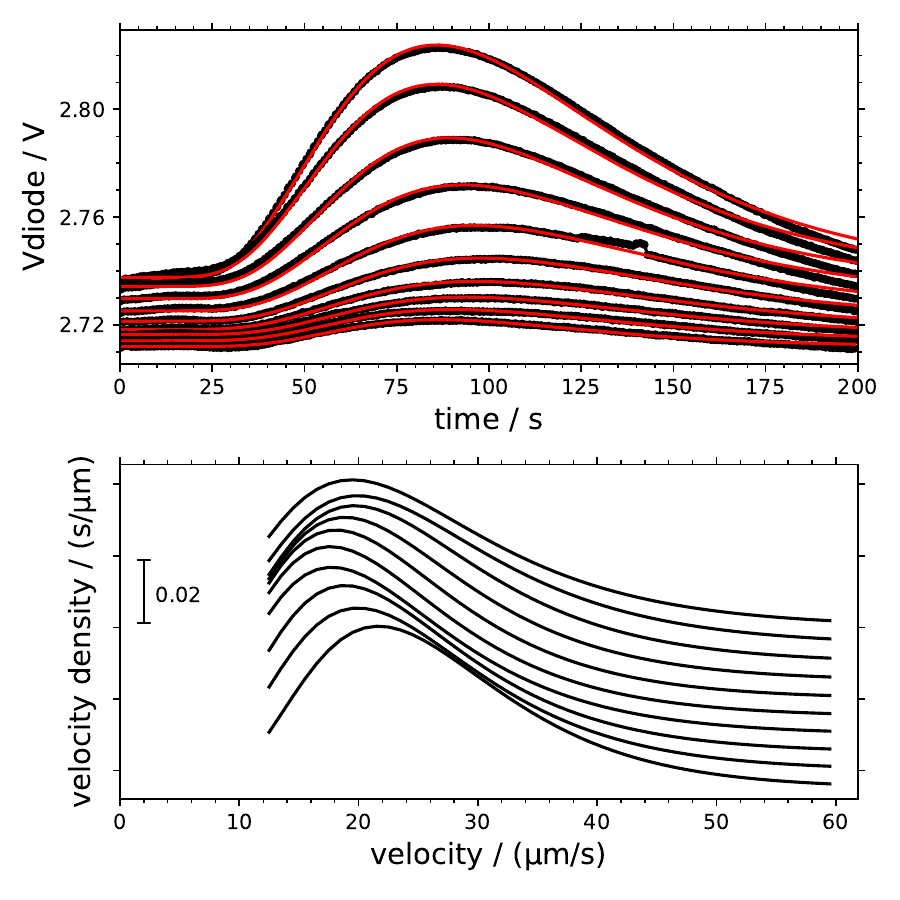}
    \caption{Photodetector output as a function of time elapsed
      since reversal of the magnetic field. After approximate
      \SI{30}{s}, the first MC-1 pass the light beam and scatter the
      light. Maximum light extinction is reached about \SI{90}{s}. The
      experiment is repeated eight times in intervals of
      \SI{440}{s} between measurements. The
      responses fit relatively well to a log-normal
      distribution (red lines). These fits can be inverted to obtain
      the velocity distribution of the MC-1 (bottom curve). For
      clarity, these curves are offset by \SI{0.005}{s/\um} from top to
      bottom. The velocity distribution remains more or less constant.}
    \label{fig:MRTFit}
\end{figure}

\begin{figure}
  \centering
    \includegraphics[width=\widefigurewidth]{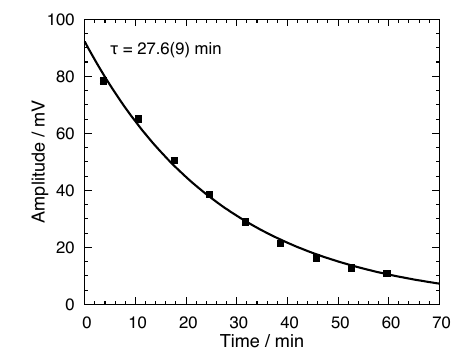}
    \caption{Decrease in amplitude of the marathon curves in
      Figure~\ref{fig:MRTFit} with increasing time agrees very well with
      an exponential decay.}
    \label{fig:Amplitude}
  \end{figure}

Figure~\ref{fig:MRTFit} shows the output of the photodetector as a
function of time elapsed after the magnetic field reversal. A series of eight
experiments are shown. For each experiment,  the cloud reaches the light path after approximately
\SI{30}{s} with a maximum density at
about \SI{90}{s}.\footnote{These experiments are performed with our
  novel amplifier. Lower intensity results in a higher detector
  voltage, see Figure~\ref{fig:VdiodeVsPLED}.} As time progresses,
the curves have a similar shape, but with a lower amplitude. Apparently,
less and less bacteria collect at the bottom of the
cuvette. The decrease in amplitude shown in
Figure~\ref{fig:Amplitude} agrees very well with an exponential decay
$\exp(-t/\tau)$ and a time constant of approximately 30~min. This suggests that  a fixed amount of bacteria is lost per
iteration. The reason for the loss is yet unclear;  this remains a topic for further investigation.

The arrival time $t$~(s) of MC-1 at the detector agrees well with a
log-normal distribution (shown as red curves in Figure~\ref{fig:MRTFit}),
\begin{equation}
  f_\text{t}(t)=\frac{1}{t\sigma\sqrt{2\pi}}
  \exp\left(\frac{-\left(\ln(t)-\mu\right)^2}{2\sigma^2}\right)\text{ (1/s)} \, ,
\end{equation}
where $\mu$ (with unit $\ln(s)$) and $\sigma$ (unitless) are the mean
and standard deviation of the natural logarithm of $t$, respectively. From the
distance of \SI{2.5}{mm} from the bottom of the cuvette to the light beam $a$, we can calculate the distribution of the velocities $v$
(m/s) as follows:
\begin{equation}
  \label{eq:speeddistribution}
  f_\text{v}(v)=\frac{1}{v \sigma\sqrt{2\pi}}
  \exp\left(\frac{-\left(\ln(\frac{a}{v})-\mu\right)^2}{2\sigma^2}\right)\text{
    (s/m)} \, .
\end{equation}
The most likely velocity, or mode of this distribution, is\footnote{Note that the most likely arrival time is $\exp(\mu-\sigma^2)$. Therefore, one
cannot simply divide the distance travelled by the most likely arrival
time to obtain the most likely velocity.}
\begin{equation}
  \label{eq:mode}
  v_\text{m} = a \exp\left(-(\mu + \sigma^2)\right) \text{ (m/s)} \, .
\end{equation}

The resulting velocity distributions are shown in the lower graph of Figure~\ref{fig:MRTFit}. The curves are offset vertically for clarity; the
top curve is the first measurement. This figure shows clearly that
the velocity distribution of the bacteria does not change
significantly with time. As the experiment duration was limited to
\SI{200}{s}, the minimum velocity that can be determined is
\SI{12.5}{\um/s}. The most likely velocity is on the order of
\SI{20}{\um/s}, and the fastest bacteria swim a rate of approximately \SI{80}{\um/s}.

Figure~\ref{fig:MRTFit} is a typical example; we have measured both faster
and slower average velocities. The measured velocity is 
considerably lower than that observed by Lef\`evre and colleagues using
high-speed microscopy~\cite{Lefevre2014}. We noted from experiments with
microfluidic chips that the velocity distribution is
strongly dependent on the duration the MC-1 have been growing in the
semi-solid medium, temperature (both too high and too low reduced
velocity) and oxygen concentration. Further experiments are required to
determine the relation between the velocity and these environmental
conditions.

\section{Discussion}
\label{sec:discussion}
Our novel MagOD magnetic absorbance instrument has proved to be a versatile
system that can be successfully applied to research on
magnetotactic bacteria. All designs and source codes have been made
available online, so that the system can be easily replicated, modified and
improved. The data presented in this paper is intended to serve as a benchmark for
future systems. We hope our efforts will inspire colleagues to improve
and apply the MagOD system for their own research. In the following, we address
possible improvements and suggestions for further research.

\subsection{Possible improvements}
There are a few issues regarding the measurement head, measurement board and software that deserve attention for future iterations
of the system. 

\subsubsection{Measurement head}

The response of the photodiode to the blue LED is nonlinear, see
Figure~\ref{fig:VdiodeVsPLED}, for which we do not have a satisfactory
explanation. Furthermore, the fact that the signal is inversely
proportional to the light intensity is counterintuitive. It may be
possible to redesign the amplifier while  maintaining the
required footprint. The noise floor in the current design is still a
factor of 30 above the theoretical limit, and there are signs that
\SI{50}{Hz} crosstalk  deteriorates the signal, see
Figure~\ref{fig:noise}. One might consider moving the AD converter
from the measurement board to the measurement head  to better protect the signal
from interference. To compensate for drift,
automatic offset correction can be applied by modulating the LED
intensity periodically.

One might consider moving the LED drivers to the measurement
head, so that the high-frequency PWM signal does not have to be
transported over the HDMI  cable. This would free up ports on the
ESP32. For this purpose, RGB LED drivers that communicate over I$^2$C are
readily available. Care should be taken that the I$^2$C clock signal
does not interfere with the detection electronics.

If suspensions with higher densities are to be observed, one might
consider using solid-state lasers that offer at least \SI{100}{} times
higher light intensity.

In contrast to commercial absorbance meters, our MagOD system does not
have a piezoelectric actuator to disperse the suspension prior to
measurement. One might  consider integrating such a component. Alternatively, one might make use of the existing coils and
attempt a voice coil actuation principle using a soft magnetic
element, an additional small coil or even a small permanent magnet
mounted such that it does not interfere with the field.

Since we use air coils, the magnetic field is simply proportional to
the current, which is measured. This works very well for fields in the
order of a few mT. For fields below \SI{1}{mT}, the magnetic
background field (for instance the Earth magnetic field) becomes
noticeable. One simple solution is to determine the background field
at the location of the setup using a three-axis Hall sensor and
compensate in a feed forward manner. However, the compensation needs
to be recalibrated every time the setup is moved. A much more precise
solution, which also corrects for soft magnetic and hard magnetic
parts of the setup, was described in detail in ref~\cite{Ahlers2022}
and is based on the determination of the magnetic response function of
the setup at the predicted detection location in the sample. Since
three-axis Hall sensors are routinely used in mobile phone mobile
phones, there have become very small and inexpensive. Therefore they
can be relatively easily positioned around the detection area. The
approach by Ahlers, in combination with a feedback loop, would thus
allow for very precise, realtime compensation in the \SI{}{\micro T}
range.

\subsubsection{Measurement board}
In future designs of the measurement board, a number of improvements
could be made as well. Even though the AD converter is capable of data
acquisition at \SI{860}{samples/s}, we achieve only  \SI{323}{samples/s}
in practice. We assume this discrepancy is caused by communication
overhead that could  be optimized.

The current implementation of the  measurement circuits 
allows only for positive currents. Modifying the circuits to allow for
bi-directional currents is straightforward, for instance by applying
an INA266 integrated current monitor.

Finally, the small form factor of micro-SD cards poses a problem in
biolab environments because they are easily lost. Removal of the SD card
can be avoided if WiFi access is available, but a USB stick may be a
better option.

\subsubsection{Software}
We expect that the software of our MagOD system will undergo the most development. In addition
to improvements of the user interface,  the main restriction
is currently that measurement recipes are based only on feed-forward instructions, i.e.\ 
iterations of a specified amount of time, field settings and LED
color. There is currently no capability to react to changes in the detected
signal. For instance, it would be very useful if the LED intensity
could be adjusted automatically to the absorbance of the suspension
under investigation. In marathon experiments, it would be convenient
if the field reversal took place at a fixed delay after the occurrence of
the peak. The current recipe language definition is not capable of
handling this type of feedback. We suspect a complete redesign of the
software is required, taking full advantage of the EPS32
capabilities. This would be a very interesting task for a (software)
engineering student.

\subsection{Possible future applications}
The four experiments we presented are merely a selection of the many possibilities offered by our
 novel MagOD system. Even without additional modification,
there are numerous possible experiments to inspire
future work.

\subsubsection{Flagellar motion}
As the MagOD system has precise field control, it allows a
simple study of the relation between field strength and
$C_\text{mag}$. It would be interesting to check whether the swimming
activity of the bacteria themselves contributes to their random
motion, which effectively would increase $kT$ and could explain the
observed difference. For instance, it would be sufficient to measure
$C_\text{mag}$ as a function of field before and after killing the
MSR-1 (for example by exposure to intense UV light or formaldehyde).

\subsubsection{Multi-color optical density}
So far, we have measured the transmission through MSR-1 cultures only
under green light. However, we noticed that the color of cultures
changes with elapsed time. We speculate that these color changes may
be caused by an increase in bacteria size and/or formation of
magnetosomes. For long-term analysis as illustrated in Figure~\ref{fig:OD}, it may
therefore be useful to measure this at different wavelengths. The MagOD
system can easily achieve this by measuring iteratively with 
red, green and blue LEDs. Multiple wavelengths may be combined with the
addition of an indicator agent that changes its absorbance spectrum based on changed conditions.

An example of such an indicator is Resazurin, which reacts to an increase
in oxygen concentration with a shift in the absorbance spectrum toward
 red. The ratio between the absorbance in the red and green
channels could therefore be a measure of the oxygen concentration in
the culture, using the blue channel for calibration.

\subsubsection{Modulated light intensity}
The intensity of the LEDs can be varied rapidly, as illustrated in
Figure~\ref{fig:ModulatedSignal860}. One can use this modulation for
differential measurements to correct for interference signals due to
changes in environmental light or crosstalk on the analog signal
wiring.

Modulation of the light intensity would also provide information about
the photosensitivity of the bacteria. For instance, one could  measure
$C_\text{mag}$ in the red channel before and after a pulse with
intense blue light.

\subsubsection{Combined marathon and $C_\text{mag}$ }
We demonstrated $C_\text{mag}$ measurements on MSR-1 bacteria and
marathon tests on MC-1 bacteria. It is straightforward to combine the
marathon test with $C_\text{mag}$ measurements. The vertical field
($z$-direction) should then be switched between zero and, for instance,
a positive value, whereas the field along the light path
($x$-direction) should be switched between zero and alternatively
positive and negative values, hence $(x,z)={ (0,1), (1,0), (0,1),
  (-1,0)}$, etc. Such an experiment might reveal whether the velocity distribution  is related to a distribution of magnetosome strength as well.

\subsubsection{Sedimentation}
We often observed an initial increase of light transmission after
loading a sample with bacteria. We therefore usually waited until the
signal settled. However, there may be information to be extracted from this behavior. We
suspect the increase in transmission is caused by sedimentation of
debris, such as dead bacteria. If the dead bacteria contain magnetosomes,
they will still rotate in the magnetic field. Therefore a measurement of
$C_\text{mag}$ during sedimentation might provide additional information
about the status of the culture.

Moreover, it is very simple to drive only one coil of the vertical
coil set. In this way, one can generate field gradients that would pull
magnetic debris either up or down, thus decelerating or accelerating
the sedimentation process.

\subsubsection{Suspensions of magnetic nanoparticles}
We often use our MagOD system with a suspension of magnetic micro- and
nanoparticles. This works particularly well for magnetic
needles~\cite{Song2021} and magnetic discs~\cite{Loethman2018a}. In
principle, spherical particles should not show a change in
transmission under rotation of an external field. However, magnetic particles
have a tendency to form chains that align with the field, see
for instance the work by Gao~\cite{Gao2013}. Angle and field-dependent transmission measurements in the MagOD could therefore
provide information about the dynamic interaction between
particles. Use of our MagOD system could perhaps  be extended even beyond the
magnetotactic research community.



\section{Conclusion}

We constructed a magnetic spectrophotometer (magnetic optical density
meter, or MagOD) that analyzes the amount of light transmitted
through a suspension of a magnetotactic bacteria in a transparent
cuvette under application of a magnetic field.

Light transmission measurements with our novel MagOD system were
compared with those obtained with a commercial instrument (Eppendorf BioPhotoMeter) using
 a dilution series of a magnetic nanoparticle suspension. The
deviation between our MagOD system and the commercial instrument is
less than \SI{0.2}{} in terms of relative absorbance for wavelengths
ranging from \SI{460}{} to
\SI{645}{nm}\cmtr{Figure~\ref{fig:EppendorfComparison}}.  However, the blue
channel  suffers from nonlinearity and should only be used at
low intensities\cmtr{Figure~\ref{fig:VdiodeVsPLED}}. The dynamic range
(from noise level to maximum signal) of our novel MagOD system is \SI{88}{dB} (optical density of
\SI{4.4}{}), whereas the commercial system reaches \SI{53}{dB} (optical density of
\SI{2.6}{}).\cmtr{Figure~\ref{fig:noise}} In addition, our MagOD system is
considerably faster, with a sample rate of
\SI{323}{samples/s}\cmtr{Figure~\ref{fig:ModulatedSignal860}}, whereas the
commercial instrument has a sampling time in excess of
\SI{1}{\second}.

The magnetic field can be applied in three directions, with a set-point
resolution of \SI{70}{nT} and a ripple of less than \SI{50}{\micro
  T}. The maximum field is \SI{5.1(1)}{mT}, but is limited in duration due
to coil heating.\cmtr{Figure~\ref{fig:BVsICoil}} When a field of
\SI{1.0}{mT} is continuously applied, the temperature increase of the cuvette is
approximately \SI{1}{K/h} and limited to \SI{2.1(3)}{K}\cmtr{Figure~\ref{fig:TVsTime}}.


The MagOD system was used to characterize various aspects of MSR-1 and
MC-1 magnetotactic bacteria.
By means of the magnetic field, MSR-1 bacteria were oriented at different angles
with respect to the light path.  The transmission rate is high when bacteria 
are aligned along the light beam and lower when the bacteria are
aligned perpendicular to the light path. The relation between angle
and optical density can be approximated relatively well by a sine function.\cmtr{Figure~\ref{fig:gopt}}

The difference in transmission rates allows us to derive a measure for the
amount of magnetic bacteria. This amount is commonly expressed as a
ratio $C_\text{mag}$, which is a parameter that increases with the
relative fraction of magnetic bacteria compared to the total number of
bacteria. It can also be expressed as a difference
$\Delta_\text{OD}$, which is a measure for the absolute amount of
magnetic bacteria. Both parameters increase with applied field
strength\cmtr{Figure~\ref{fig:bac_response}} in a way that is  in agreement within
measurement error with a simple model based on Brownian
motion\cmtr{Figure~\ref{fig:gParPer}}.

We used our novel MagOD system to continuously monitor the development of a
culture of MSR-1 magnetotactic bacteria for \SI{5}{days}.  We
recorded the optical density $O_\text{D}$, change in light
transmission under rotation of the magnetic field $C_{\mathrm{mag}}$
and $\Delta_\text{OD}$, and the rotation velocity of the bacteria
$\gamma$.  We were able to distinguish clearly the separate growth phases (lag,
exponential growth, and stationary). The increase in magnetic response
$C_{\mathrm{mag}}$ and $\Delta_\text{OD}$ takes place during the stationary
phase.\cmtr{Figure~\ref{fig:OD}}

Unipolar bacteria such as MC-1 can be collected at the bottom of the
cuvette with a vertical magnetic field. Upon reversal of the field, the
entire group departs from the bottom and will arrive at the light
beam, causing a dip in the transmitted light. This ``marathon'' test
allows us to measure the velocity distribution.

The arrival times can be accurately described by a log-normal
distribution, with a mode (most frequently occurring velocity) of
\SI{20}{\um/s}. The maximum velocity we observed is on the order of
\SI{80}{\um/s}.\cmtr{Figure~\ref{fig:MRTFit}} The amount of bacteria
participating in the marathon test decreases exponentially with each
test with a time constant of approximately 30~min\cmtr{Figure~\ref{fig:Amplitude} }.

The dedicated magnetic optical density meter presented here is
relatively simple and inexpensive, yet the data that can be extracted
from magnetotactic bacteria cultures is rich in detail.  All
information for the construction of the device, including 3D print
designs, printed circuit board layouts and code for the microprocessor,
has been made available online. The authors trust that the
magnetotactic bacteria community will benefit from our work, and that
the MagOD instrument will become a valuable tool for research in this field.


\section*{Supplementary Online Material}
The online supplementary material contains a python script
(\texttt{angular.py}) to numerically integrate the equations in
section~\ref{sec:brownian}, and a video of the LED intensity
(\texttt{MagODLEDProjection.mov}) accompanying
figure~\ref{fig:LEDProjection}.

\section*{Acknowledgments}
The authors thank Christopher Lef\`evre for teaching us how to grow
MSR-1 and MC-1 bacteria and for discussions and feedback on the
manuscript, Annissa Dieudonne for testing the first version of the
MagOD meter and for valuable feedback on the manuscript, Andreas Manz
for beneficial discussions and guidance, Prof.\ Long Fei Wu and Stefan
Klumpp for their excellent feedback on the manuscript, and Sander
Smits for 3D prints and laser cutting.

This research was funded in part by KIST Europe, Basic Research
Project 12202 (L.A., M.P., T.H., A.M., N.K) and by BBSRC, grant
BB/V010603/1 (A.F.C).

\clearpage
\appendix
\section{$C^*_\text{mag}$ and $C_\text{mag}$ approximations}
\label{sec:appendix_cmag}
The effect of a magnetic field rotation is usually small. It is
therefore  useful to express the variation with respect to the
average intensity or absorbance\footnote{$A$=$O_\text{D}$}
\begin{eqnarray}
  I_\text{s}=\frac{I(0)+I(90)}{2}\\
  A = \log(I_\text{ref}/I_\text{s})
\end{eqnarray}
by a small deviation $\alpha$
\begin{eqnarray}
  A(0)=(1+\alpha)A\\
  A(90)=(1-\alpha)A
\end{eqnarray}
so that
\begin{eqnarray}
  \Delta_\text{A}=2\alpha A
\end{eqnarray}
and
\begin{eqnarray}
  C_\text{mag}=\frac{(1+\alpha)}{(1-\alpha)}\approx 1+ 2\alpha=
  1 + \frac{\Delta_\text{A}}{A} \,.
\end{eqnarray}
The approximation is better than \SI{5}{\percent} in terms of
$C_\text{mag}-1$ for $C_\text{mag}<1.1$.

Similarly, to estimate $C^*_\text{mag}$, we can define
\begin{equation}
  \Delta I = 2 \beta I_\text{s}
\end{equation}
so that
\begin{equation}
  C^*_\text{mag}\approx 1+ \frac{I_\text{s}}{I_\text{ref}-I_\text{s}} 2\beta=\
  1+\frac{\Delta I}{I_\text{ref}-I_\text{s}} \, .
\end{equation}
Both definitions of $C_\text{mag}$ can be related by realizing that
\begin{eqnarray}
  \frac{I(0)}{I_\text{ref}}=\left(\frac{I_\text{s}}{I_\text{ref}}\right)^{1+\alpha}\approx\
  \frac{I_\text{s}}{I_\text{ref}}\left(1+\alpha\ln(I_\text{s}/ I_\text{ref})\right)
\end{eqnarray}
and similarly for $I(90)$ with $-\alpha$, so that
\begin{equation}
  \Delta I = -2 \alpha I_\text{s} \ln(I_\text{s}/I_\text{ref}) \, .
\end{equation}
Therefore, in the approximation for $C_\text{mag}$ close to
unity, the relation between these two definitions is
\begin{eqnarray}
  \frac{C_\text{mag}-1}{C^*_\text{mag}-1}\approx\
  \frac{\Delta_\text{A}}{A}\frac{I_\text{ref}-I_\text{s}}{\Delta I}= \\
  \frac{I_\text{ref}-I_\text{s}}{I_\text{s}\ln(I_\text{ref}/I_\text{s})}=\frac{(I_\text{ref}-I_s)\log(e)}{I_\text{s}A}.
\end{eqnarray}
The definitions converge for $I_\text{s} \to I_\text{ref}$  for samples with very low
optical density.

\section{Arccotangent approximation}
\label{sec:appendix_cotan}

\begin{figure}
  \centering
    \includegraphics[width=\figurewidth]{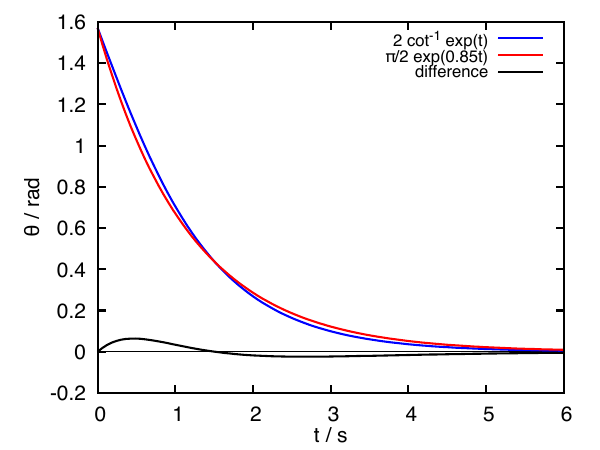}
    \caption{Approximation by an exponential function of the exact
      solution to the differential equation for the rotation of the
      bacterium as a function of time, see
      Eq.~(\ref{eq:diff_eq_solution}). The optimal fit is for a
      prefactor \SI{0.85(1)}{}, in which case the error is less
      than \SI{0.065}{rad}.}
    \label{fig:acosVsexp}
\end{figure}

For fitting purposes, the rather complicated arccotangent expression of
Eq.~(\ref{eq:diff_eq_solution}) can be approximated by a much
simpler exponential function. The fit was performed in gnuplot,
resulting in a fit parameter of \SI{0.85(1)}{}. The error is less than
\SI{0.065}{rad} as shown in Figure~\ref{fig:acosVsexp}.

\clearpage
\section{Measurements}
\label{sec:appendix_meas}
\begin{figure}
  \begin{center}
    \includegraphics[width=0.7\linewidth]{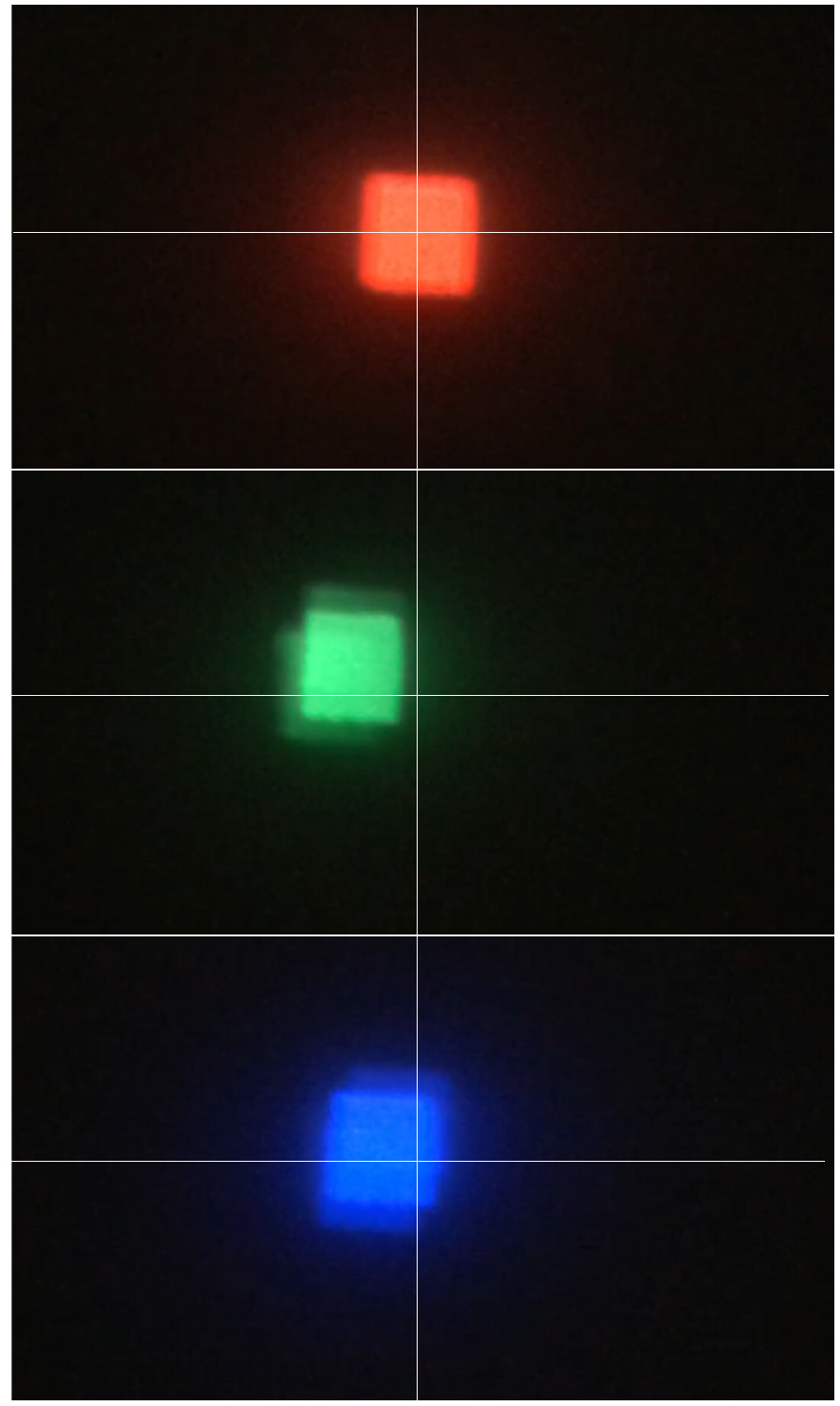}
    \caption{Projection of the light pattern of the three different
      LEDs. The \num{3}$\times$\SI{3}{mm} opening at the back of the cuvette can
      clearly be seen. The green and blue LEDs show some reflection,
      and the patterns are not aligned. The estimated misalignment is
      on the order of \SI{2}{\degree}. As the photodetector has a sensitive
      area of \num{2.7}$\times$\SI{2.7}{mm} and is  mounted directly behind the
      opening in the holder, we expect no adverse effects from
      the reflections of misalignment. A
      full video of the pattern shape as a function of  intensity is
      available in the Supplementary Material.}
    \label{fig:LEDProjection}
  \end{center}
\end{figure}

Figure~\ref{fig:LEDProjection} shows the projection onto a sheet of white paper  of the light
exiting from the measurement head (with the photodetector circuit
board removed). The images are snapshots of a video taken with an
iPhone camera for a range in LED duty cycles (measurement of Figure~\ref{fig:SignalVsPower}). The full video is available in the Supplementary
Material, see MagODLEDprojection.mov. The opening behind the cuvette is a
square hole of \num{3}$\times$\SI{3}{mm}, which is clearly visible. The photodetector
itself has an area of \num{2.7}$\times$\SI{2.7}{mm}, hence it collects the inner
portion of this pattern. For the green and blue LEDs, echo images
can be observed. The three patterns do not align, which is most likely
caused by the fact that the three LEDs in the WP154 housing are not
centered on the axis of the front lens. From the distance between the
projected image and the LED (approximately \SI{15}{cm}), with a
maximum shift of about \SI{5}{mm}, we estimate that
the misalignment is on the order of \SI{2}{\degree}.  As the
photodetector is mounted directly behind the opening behind the
cuvette, this misalignment is of no consequence.

Figure~\ref{fig:SignalVsPower} shows the intensity on the photodetector
as a function of the duty cycle of each of the three LEDs for a
cuvette filled with water and three different dilutions of a EMG304
magnetic nanoparticle suspension.  The absorbance relative to the water-filled cuvette is determined from the difference in slopes. This
measurement is used for the MagOD data points in
Figure~\ref{fig:EppendorfComparison}. The blue LED suffers from
artefacts. The slope is not constant, but lower at high
intensities. Furthermore, there is a small step at an intensity of about
\num{0.6}. For the estimate of absorbance of the blue LED, only the
linear region at low intensity was used.

\begin{figure}
  \begin{center}
    \includegraphics[width=\linewidth]{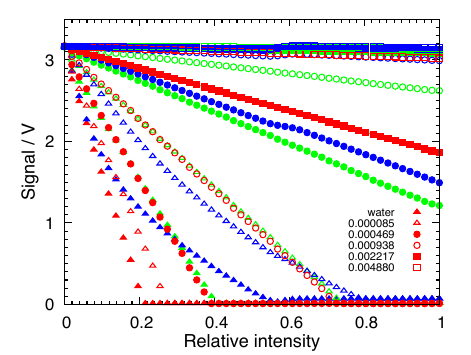}
    \caption{Signal as a function of the LED duty cycle  for
      the three different LED wavelengths as well as for a water reference and
      three different dilutions of a EMG304 magnetic nanoparticle
      suspension.}
    \label{fig:SignalVsPower}
  \end{center}
\end{figure}


\clearpage
\bibliographystyle{apsrev4-2-modified}
\bibliography{paperbase}

\begin{thebibliography}{52}%
\makeatletter
\providecommand \@ifxundefined [1]{%
 \@ifx{#1\undefined}
}%
\providecommand \@ifnum [1]{%
 \ifnum #1\expandafter \@firstoftwo
 \else \expandafter \@secondoftwo
 \fi
}%
\providecommand \@ifx [1]{%
 \ifx #1\expandafter \@firstoftwo
 \else \expandafter \@secondoftwo
 \fi
}%
\providecommand \natexlab [1]{#1}%
\providecommand \enquote  [1]{``#1''}%
\providecommand \bibnamefont  [1]{#1}%
\providecommand \bibfnamefont [1]{#1}%
\providecommand \citenamefont [1]{#1}%
\providecommand \href@noop [0]{\@secondoftwo}%
\providecommand \href [0]{\begingroup \@sanitize@url \@href}%
\providecommand \@href[1]{\@@startlink{#1}\@@href}%
\providecommand \@@href[1]{\endgroup#1\@@endlink}%
\providecommand \@sanitize@url [0]{\catcode `\\12\catcode `\$12\catcode
  `\&12\catcode `\#12\catcode `\^12\catcode `\_12\catcode `\%12\relax}%
\providecommand \@@startlink[1]{}%
\providecommand \@@endlink[0]{}%
\providecommand \url  [0]{\begingroup\@sanitize@url \@url }%
\providecommand \@url [1]{\endgroup\@href {#1}{\urlprefix }}%
\providecommand \urlprefix  [0]{URL }%
\providecommand \Eprint [0]{\href }%
\providecommand \doibase [0]{https://doi.org/}%
\providecommand \selectlanguage [0]{\@gobble}%
\providecommand \bibinfo  [0]{\@secondoftwo}%
\providecommand \bibfield  [0]{\@secondoftwo}%
\providecommand \translation [1]{[#1]}%
\providecommand \BibitemOpen [0]{}%
\providecommand \bibitemStop [0]{}%
\providecommand \bibitemNoStop [0]{.\EOS\space}%
\providecommand \EOS [0]{\spacefactor3000\relax}%
\providecommand \BibitemShut  [1]{\csname bibitem#1\endcsname}%
\let\auto@bib@innerbib\@empty
\bibitem [{\citenamefont {Frankel}\ \emph {et~al.}(1979)\citenamefont
  {Frankel}, \citenamefont {Blakemore},\ and\ \citenamefont
  {Wolfe}}]{Frankel1979}%
  \BibitemOpen
  \bibfield  {author} {\bibinfo {author} {\bibfnamefont {R.~B.}\ \bibnamefont
  {Frankel}}, \bibinfo {author} {\bibfnamefont {R.~P.}\ \bibnamefont
  {Blakemore}},\ and\ \bibinfo {author} {\bibfnamefont {R.~S.}\ \bibnamefont
  {Wolfe}},\ }\href {http://www.jstor.org/stable/1747327} {\bibfield  {journal}
  {\bibinfo  {journal} {Science}\ }\textbf {\bibinfo {volume} {203}},\ \bibinfo
  {pages} {1355} (\bibinfo {year} {1979})}\BibitemShut {NoStop}%
\bibitem [{\citenamefont {Farina}\ \emph {et~al.}(1990)\citenamefont {Farina},
  \citenamefont {Esquivel},\ and\ \citenamefont {De~Barros}}]{Farina1990}%
  \BibitemOpen
  \bibfield  {author} {\bibinfo {author} {\bibfnamefont {M.}~\bibnamefont
  {Farina}}, \bibinfo {author} {\bibfnamefont {D.}~\bibnamefont {Esquivel}},\
  and\ \bibinfo {author} {\bibfnamefont {H.}~\bibnamefont {De~Barros}},\ }\href
  {https://www.scopus.com/inward/record.uri?eid=2-s2.0-0025197439&partnerID=40&md5=0ce3d8b9bc4236f74f80e17e9bafdb4f}
  {\bibfield  {journal} {\bibinfo  {journal} {Nature}\ }\textbf {\bibinfo
  {volume} {343}},\ \bibinfo {pages} {256} (\bibinfo {year}
  {1990})}\BibitemShut {NoStop}%
\bibitem [{\citenamefont {Lef\`evre}\ \emph {et~al.}(2014)\citenamefont
  {Lef\`evre}, \citenamefont {Bennet}, \citenamefont {Landau}, \citenamefont
  {Vach}, \citenamefont {Pignol}, \citenamefont {Bazylinski}, \citenamefont
  {Frankel}, \citenamefont {Klumpp},\ and\ \citenamefont
  {Faivre}}]{Lefevre2014}%
  \BibitemOpen
  \bibfield  {author} {\bibinfo {author} {\bibfnamefont {C.}~\bibnamefont
  {Lef\`evre}}, \bibinfo {author} {\bibfnamefont {M.}~\bibnamefont {Bennet}},
  \bibinfo {author} {\bibfnamefont {L.}~\bibnamefont {Landau}}, \bibinfo
  {author} {\bibfnamefont {P.}~\bibnamefont {Vach}}, \bibinfo {author}
  {\bibfnamefont {D.}~\bibnamefont {Pignol}}, \bibinfo {author} {\bibfnamefont
  {D.}~\bibnamefont {Bazylinski}}, \bibinfo {author} {\bibfnamefont
  {R.}~\bibnamefont {Frankel}}, \bibinfo {author} {\bibfnamefont
  {S.}~\bibnamefont {Klumpp}},\ and\ \bibinfo {author} {\bibfnamefont
  {D.}~\bibnamefont {Faivre}},\ }\href
  {https://doi.org/10.1016/j.bpj.2014.05.043} {\bibfield  {journal} {\bibinfo
  {journal} {Biophysical Journal}\ }\textbf {\bibinfo {volume} {107}},\
  \bibinfo {pages} {527} (\bibinfo {year} {2014})}\BibitemShut {NoStop}%
\bibitem [{\citenamefont {Bellini}(2009)}]{Bellini2009}%
  \BibitemOpen
  \bibfield  {author} {\bibinfo {author} {\bibfnamefont {S.}~\bibnamefont
  {Bellini}},\ }\href {https://doi.org/10.1007/s00343-009-0003-5} {\bibfield
  {journal} {\bibinfo  {journal} {Chinese Journal of Oceanology and Limnology}\
  }\textbf {\bibinfo {volume} {27}},\ \bibinfo {pages} {3} (\bibinfo {year}
  {2009})}\BibitemShut {NoStop}%
\bibitem [{\citenamefont {Blakemore}\ \emph {et~al.}(1979)\citenamefont
  {Blakemore}, \citenamefont {Maratea},\ and\ \citenamefont
  {Wolfe}}]{Blakemore1979}%
  \BibitemOpen
  \bibfield  {author} {\bibinfo {author} {\bibfnamefont {R.~P.}\ \bibnamefont
  {Blakemore}}, \bibinfo {author} {\bibfnamefont {D.}~\bibnamefont {Maratea}},\
  and\ \bibinfo {author} {\bibfnamefont {R.~S.}\ \bibnamefont {Wolfe}},\ }\href
  {http://jb.asm.org/content/140/2/720.abstract} {\bibfield  {journal}
  {\bibinfo  {journal} {Journal of Bacteriology}\ }\textbf {\bibinfo {volume}
  {140}},\ \bibinfo {pages} {720} (\bibinfo {year} {1979})}\BibitemShut
  {NoStop}%
\bibitem [{\citenamefont {Lee}\ \emph {et~al.}(2004)\citenamefont {Lee},
  \citenamefont {Purdon}, \citenamefont {Chu},\ and\ \citenamefont
  {Westervelt}}]{Lee2004b}%
  \BibitemOpen
  \bibfield  {author} {\bibinfo {author} {\bibfnamefont {H.}~\bibnamefont
  {Lee}}, \bibinfo {author} {\bibfnamefont {A.}~\bibnamefont {Purdon}},
  \bibinfo {author} {\bibfnamefont {V.}~\bibnamefont {Chu}},\ and\ \bibinfo
  {author} {\bibfnamefont {R.}~\bibnamefont {Westervelt}},\ }\href
  {https://doi.org/10.1021/nl049562x} {\bibfield  {journal} {\bibinfo
  {journal} {Nano Letters}\ }\textbf {\bibinfo {volume} {4}},\ \bibinfo {pages}
  {995} (\bibinfo {year} {2004})}\BibitemShut {NoStop}%
\bibitem [{\citenamefont {{Khalil}}\ \emph {et~al.}(2013)\citenamefont
  {{Khalil}}, \citenamefont {{Pichel}}, \citenamefont {{Abelmann}},\ and\
  \citenamefont {{Misra}}}]{Khalil2013b}%
  \BibitemOpen
  \bibfield  {author} {\bibinfo {author} {\bibfnamefont {I.~S.~M.}\
  \bibnamefont {{Khalil}}}, \bibinfo {author} {\bibfnamefont {M.~P.}\
  \bibnamefont {{Pichel}}}, \bibinfo {author} {\bibfnamefont {L.}~\bibnamefont
  {{Abelmann}}},\ and\ \bibinfo {author} {\bibfnamefont {S.}~\bibnamefont
  {{Misra}}},\ }\href {https://doi.org/10.1177/0278364913479412} {\bibfield
  {journal} {\bibinfo  {journal} {The International Journal of Robotics
  Research}\ }\textbf {\bibinfo {volume} {32}},\ \bibinfo {pages} {637}
  (\bibinfo {year} {2013})}\BibitemShut {NoStop}%
\bibitem [{\citenamefont {Pichel}\ \emph {et~al.}(2018)\citenamefont {Pichel},
  \citenamefont {Hageman}, \citenamefont {Khalil}, \citenamefont {Manz},\ and\
  \citenamefont {Abelmann}}]{Pichel2018}%
  \BibitemOpen
  \bibfield  {author} {\bibinfo {author} {\bibfnamefont {M.~P.}\ \bibnamefont
  {Pichel}}, \bibinfo {author} {\bibfnamefont {T.~A.~G.}\ \bibnamefont
  {Hageman}}, \bibinfo {author} {\bibfnamefont {I.~S.~M.}\ \bibnamefont
  {Khalil}}, \bibinfo {author} {\bibfnamefont {A.}~\bibnamefont {Manz}},\ and\
  \bibinfo {author} {\bibfnamefont {L.}~\bibnamefont {Abelmann}},\ }\href
  {https://doi.org/10.1016/j.jmmm.2018.04.004} {\bibfield  {journal} {\bibinfo
  {journal} {Journal of Magnetism and Magnetic Materials}\ }\textbf {\bibinfo
  {volume} {460}},\ \bibinfo {pages} {340} (\bibinfo {year}
  {2018})}\BibitemShut {NoStop}%
\bibitem [{\citenamefont {Futschik}\ \emph {et~al.}(1989)\citenamefont
  {Futschik}, \citenamefont {Pfützner}, \citenamefont {Doblander},
  \citenamefont {Schönhuber}, \citenamefont {Dobeneck}, \citenamefont
  {Petersen},\ and\ \citenamefont {Vali}}]{Futschik1989}%
  \BibitemOpen
  \bibfield  {author} {\bibinfo {author} {\bibfnamefont {K.}~\bibnamefont
  {Futschik}}, \bibinfo {author} {\bibfnamefont {H.}~\bibnamefont {Pfützner}},
  \bibinfo {author} {\bibfnamefont {A.}~\bibnamefont {Doblander}}, \bibinfo
  {author} {\bibfnamefont {P.}~\bibnamefont {Schönhuber}}, \bibinfo {author}
  {\bibfnamefont {T.}~\bibnamefont {Dobeneck}}, \bibinfo {author}
  {\bibfnamefont {N.}~\bibnamefont {Petersen}},\ and\ \bibinfo {author}
  {\bibfnamefont {H.}~\bibnamefont {Vali}},\ }\href
  {https://doi.org/10.1088/0031-8949/40/4/016} {\bibfield  {journal} {\bibinfo
  {journal} {Physica Scripta}\ }\textbf {\bibinfo {volume} {40}},\ \bibinfo
  {pages} {518} (\bibinfo {year} {1989})}\BibitemShut {NoStop}%
\bibitem [{\citenamefont {Harasko}\ \emph {et~al.}(1995)\citenamefont
  {Harasko}, \citenamefont {Pfutzner},\ and\ \citenamefont
  {Futschik}}]{Harasko1995}%
  \BibitemOpen
  \bibfield  {author} {\bibinfo {author} {\bibfnamefont {G.}~\bibnamefont
  {Harasko}}, \bibinfo {author} {\bibfnamefont {H.}~\bibnamefont {Pfutzner}},\
  and\ \bibinfo {author} {\bibfnamefont {K.}~\bibnamefont {Futschik}},\ }\href
  {https://doi.org/10.1109/20.364766} {\bibfield  {journal} {\bibinfo
  {journal} {{IEEE} Transactions on Magnetics}\ }\textbf {\bibinfo {volume}
  {31}},\ \bibinfo {pages} {938} (\bibinfo {year} {1995})}\BibitemShut
  {NoStop}%
\bibitem [{\citenamefont {Hergt}\ \emph {et~al.}(2005)\citenamefont {Hergt},
  \citenamefont {Hiergeist}, \citenamefont {Zeisberger}, \citenamefont
  {Schüler}, \citenamefont {Heyen}, \citenamefont {Hilger},\ and\
  \citenamefont {Kaiser}}]{Hergt2005}%
  \BibitemOpen
  \bibfield  {author} {\bibinfo {author} {\bibfnamefont {R.}~\bibnamefont
  {Hergt}}, \bibinfo {author} {\bibfnamefont {R.}~\bibnamefont {Hiergeist}},
  \bibinfo {author} {\bibfnamefont {M.}~\bibnamefont {Zeisberger}}, \bibinfo
  {author} {\bibfnamefont {D.}~\bibnamefont {Schüler}}, \bibinfo {author}
  {\bibfnamefont {U.}~\bibnamefont {Heyen}}, \bibinfo {author} {\bibfnamefont
  {I.}~\bibnamefont {Hilger}},\ and\ \bibinfo {author} {\bibfnamefont {W.~A.}\
  \bibnamefont {Kaiser}},\ }\href {https://doi.org/10.1016/j.jmmm.2005.01.047}
  {\bibfield  {journal} {\bibinfo  {journal} {Journal of Magnetism and Magnetic
  Materials}\ }\textbf {\bibinfo {volume} {293}},\ \bibinfo {pages} {80}
  (\bibinfo {year} {2005})}\BibitemShut {NoStop}%
\bibitem [{\citenamefont {Gandia}\ \emph {et~al.}(2019)\citenamefont {Gandia},
  \citenamefont {Gandarias}, \citenamefont {Rodrigo}, \citenamefont
  {Robles-Garc{\'{\i}}a}, \citenamefont {Das}, \citenamefont {Garaio},
  \citenamefont {Garc{\'{\i}}a}, \citenamefont {Phan}, \citenamefont
  {Srikanth}, \citenamefont {Orue}, \citenamefont {Alonso}, \citenamefont
  {Muela},\ and\ \citenamefont {Fdez-Gubieda}}]{Gandia2019}%
  \BibitemOpen
  \bibfield  {author} {\bibinfo {author} {\bibfnamefont {D.}~\bibnamefont
  {Gandia}}, \bibinfo {author} {\bibfnamefont {L.}~\bibnamefont {Gandarias}},
  \bibinfo {author} {\bibfnamefont {I.}~\bibnamefont {Rodrigo}}, \bibinfo
  {author} {\bibfnamefont {J.}~\bibnamefont {Robles-Garc{\'{\i}}a}}, \bibinfo
  {author} {\bibfnamefont {R.}~\bibnamefont {Das}}, \bibinfo {author}
  {\bibfnamefont {E.}~\bibnamefont {Garaio}}, \bibinfo {author} {\bibfnamefont
  {J.~{\'{A}}.}\ \bibnamefont {Garc{\'{\i}}a}}, \bibinfo {author}
  {\bibfnamefont {M.-H.}\ \bibnamefont {Phan}}, \bibinfo {author}
  {\bibfnamefont {H.}~\bibnamefont {Srikanth}}, \bibinfo {author}
  {\bibfnamefont {I.}~\bibnamefont {Orue}}, \bibinfo {author} {\bibfnamefont
  {J.}~\bibnamefont {Alonso}}, \bibinfo {author} {\bibfnamefont
  {A.}~\bibnamefont {Muela}},\ and\ \bibinfo {author} {\bibfnamefont {M.~L.}\
  \bibnamefont {Fdez-Gubieda}},\ }\href
  {https://doi.org/10.1002/smll.201902626} {\bibfield  {journal} {\bibinfo
  {journal} {Small}\ }\textbf {\bibinfo {volume} {15}},\ \bibinfo {pages}
  {1902626} (\bibinfo {year} {2019})}\BibitemShut {NoStop}%
\bibitem [{\citenamefont {Makela}\ \emph {et~al.}(2022)\citenamefont {Makela},
  \citenamefont {Schott}, \citenamefont {Madsen}, \citenamefont {Greeson},\
  and\ \citenamefont {Contag}}]{Makela2022}%
  \BibitemOpen
  \bibfield  {author} {\bibinfo {author} {\bibfnamefont {A.~V.}\ \bibnamefont
  {Makela}}, \bibinfo {author} {\bibfnamefont {M.~A.}\ \bibnamefont {Schott}},
  \bibinfo {author} {\bibfnamefont {C.~S.}\ \bibnamefont {Madsen}}, \bibinfo
  {author} {\bibfnamefont {E.~M.}\ \bibnamefont {Greeson}},\ and\ \bibinfo
  {author} {\bibfnamefont {C.~H.}\ \bibnamefont {Contag}},\ }\href
  {https://doi.org/10.1021/acs.nanolett.1c05042} {\bibfield  {journal}
  {\bibinfo  {journal} {Nano Letters}\ }\textbf {\bibinfo {volume} {22}},\
  \bibinfo {pages} {4630} (\bibinfo {year} {2022})}\BibitemShut {NoStop}%
\bibitem [{\citenamefont {Martel}\ and\ \citenamefont
  {Mohammadi}(2010)}]{Martel2010}%
  \BibitemOpen
  \bibfield  {author} {\bibinfo {author} {\bibfnamefont {S.}~\bibnamefont
  {Martel}}\ and\ \bibinfo {author} {\bibfnamefont {M.}~\bibnamefont
  {Mohammadi}},\ }\href {https://doi.org/10.1109/ROBOT.2010.5509752} {\bibfield
   {journal} {\bibinfo  {journal} {Proceedings - IEEE International Conference
  on Robotics and Automation}\ ,\ \bibinfo {pages} {500}} (\bibinfo {year}
  {2010})}\BibitemShut {NoStop}%
\bibitem [{\citenamefont {Martel}\ \emph {et~al.}(2009)\citenamefont {Martel},
  \citenamefont {Mohammadi}, \citenamefont {Felfoul}, \citenamefont {Lu},\ and\
  \citenamefont {Pouponneau}}]{Martel2009}%
  \BibitemOpen
  \bibfield  {author} {\bibinfo {author} {\bibfnamefont {S.}~\bibnamefont
  {Martel}}, \bibinfo {author} {\bibfnamefont {M.}~\bibnamefont {Mohammadi}},
  \bibinfo {author} {\bibfnamefont {O.}~\bibnamefont {Felfoul}}, \bibinfo
  {author} {\bibfnamefont {Z.}~\bibnamefont {Lu}},\ and\ \bibinfo {author}
  {\bibfnamefont {P.}~\bibnamefont {Pouponneau}},\ }\href
  {https://doi.org/10.1177/0278364908100924} {\bibfield  {journal} {\bibinfo
  {journal} {International Journal of Robotics Research}\ }\textbf {\bibinfo
  {volume} {28}},\ \bibinfo {pages} {571} (\bibinfo {year} {2009})}\BibitemShut
  {NoStop}%
\bibitem [{\citenamefont {Khalil}\ \emph {et~al.}(2013)\citenamefont {Khalil},
  \citenamefont {Magdanz}, \citenamefont {Sanchez}, \citenamefont {Schmidt},
  \citenamefont {Abelmann},\ and\ \citenamefont {Misra}}]{Khalil2013}%
  \BibitemOpen
  \bibfield  {author} {\bibinfo {author} {\bibfnamefont {I.}~\bibnamefont
  {Khalil}}, \bibinfo {author} {\bibfnamefont {V.}~\bibnamefont {Magdanz}},
  \bibinfo {author} {\bibfnamefont {S.}~\bibnamefont {Sanchez}}, \bibinfo
  {author} {\bibfnamefont {O.}~\bibnamefont {Schmidt}}, \bibinfo {author}
  {\bibfnamefont {L.}~\bibnamefont {Abelmann}},\ and\ \bibinfo {author}
  {\bibfnamefont {S.}~\bibnamefont {Misra}},\ }\href
  {https://doi.org/10.1109/EMBC.2013.6610745} {\bibfield  {journal} {\bibinfo
  {journal} {Proceedings of the Annual International Conference of the IEEE
  Engineering in Medicine and Biology Society, EMBS}\ ,\ \bibinfo {pages}
  {5299}} (\bibinfo {year} {2013})}\BibitemShut {NoStop}%
\bibitem [{\citenamefont {De~Lanauze}\ \emph {et~al.}(2014)\citenamefont
  {De~Lanauze}, \citenamefont {Felfoul}, \citenamefont {Turcot}, \citenamefont
  {Mohammadi},\ and\ \citenamefont {Martel}}]{DeLanauze2014}%
  \BibitemOpen
  \bibfield  {author} {\bibinfo {author} {\bibfnamefont {D.}~\bibnamefont
  {De~Lanauze}}, \bibinfo {author} {\bibfnamefont {O.}~\bibnamefont {Felfoul}},
  \bibinfo {author} {\bibfnamefont {J.-P.}\ \bibnamefont {Turcot}}, \bibinfo
  {author} {\bibfnamefont {M.}~\bibnamefont {Mohammadi}},\ and\ \bibinfo
  {author} {\bibfnamefont {S.}~\bibnamefont {Martel}},\ }\href
  {https://doi.org/10.1177/0278364913500543} {\bibfield  {journal} {\bibinfo
  {journal} {International Journal of Robotics Research}\ }\textbf {\bibinfo
  {volume} {33}},\ \bibinfo {pages} {359} (\bibinfo {year} {2014})}\BibitemShut
  {NoStop}%
\bibitem [{\citenamefont {Felfoul}\ \emph {et~al.}(2016)\citenamefont
  {Felfoul}, \citenamefont {Mohammadi}, \citenamefont {Taherkhani},
  \citenamefont {De~Lanauze}, \citenamefont {Zhong~Xu}, \citenamefont {Loghin},
  \citenamefont {Essa}, \citenamefont {Jancik}, \citenamefont {Houle},
  \citenamefont {Lafleur}, \citenamefont {Gaboury}, \citenamefont {Tabrizian},
  \citenamefont {Kaou}, \citenamefont {Atkin}, \citenamefont {Vuong},
  \citenamefont {Batist}, \citenamefont {Beauchemin}, \citenamefont
  {Radzioch},\ and\ \citenamefont {Martel}}]{Felfoul2016}%
  \BibitemOpen
  \bibfield  {author} {\bibinfo {author} {\bibfnamefont {O.}~\bibnamefont
  {Felfoul}}, \bibinfo {author} {\bibfnamefont {M.}~\bibnamefont {Mohammadi}},
  \bibinfo {author} {\bibfnamefont {S.}~\bibnamefont {Taherkhani}}, \bibinfo
  {author} {\bibfnamefont {D.}~\bibnamefont {De~Lanauze}}, \bibinfo {author}
  {\bibfnamefont {Y.}~\bibnamefont {Zhong~Xu}}, \bibinfo {author}
  {\bibfnamefont {D.}~\bibnamefont {Loghin}}, \bibinfo {author} {\bibfnamefont
  {S.}~\bibnamefont {Essa}}, \bibinfo {author} {\bibfnamefont {S.}~\bibnamefont
  {Jancik}}, \bibinfo {author} {\bibfnamefont {D.}~\bibnamefont {Houle}},
  \bibinfo {author} {\bibfnamefont {M.}~\bibnamefont {Lafleur}}, \bibinfo
  {author} {\bibfnamefont {L.}~\bibnamefont {Gaboury}}, \bibinfo {author}
  {\bibfnamefont {M.}~\bibnamefont {Tabrizian}}, \bibinfo {author}
  {\bibfnamefont {N.}~\bibnamefont {Kaou}}, \bibinfo {author} {\bibfnamefont
  {M.}~\bibnamefont {Atkin}}, \bibinfo {author} {\bibfnamefont
  {T.}~\bibnamefont {Vuong}}, \bibinfo {author} {\bibfnamefont
  {G.}~\bibnamefont {Batist}}, \bibinfo {author} {\bibfnamefont
  {N.}~\bibnamefont {Beauchemin}}, \bibinfo {author} {\bibfnamefont
  {D.}~\bibnamefont {Radzioch}},\ and\ \bibinfo {author} {\bibfnamefont
  {S.}~\bibnamefont {Martel}},\ }\href {https://doi.org/10.1038/nnano.2016.137}
  {\bibfield  {journal} {\bibinfo  {journal} {Nature Nanotechnology}\ }\textbf
  {\bibinfo {volume} {11}},\ \bibinfo {pages} {941} (\bibinfo {year}
  {2016})}\BibitemShut {NoStop}%
\bibitem [{\citenamefont {Zingsem}\ \emph {et~al.}(2019)\citenamefont
  {Zingsem}, \citenamefont {Feggeler}, \citenamefont {Terwey}, \citenamefont
  {Ghaisari}, \citenamefont {Spoddig}, \citenamefont {Faivre}, \citenamefont
  {Meckenstock}, \citenamefont {Farle},\ and\ \citenamefont
  {Winklhofer}}]{Zingsem2019}%
  \BibitemOpen
  \bibfield  {author} {\bibinfo {author} {\bibfnamefont {B.~W.}\ \bibnamefont
  {Zingsem}}, \bibinfo {author} {\bibfnamefont {T.}~\bibnamefont {Feggeler}},
  \bibinfo {author} {\bibfnamefont {A.}~\bibnamefont {Terwey}}, \bibinfo
  {author} {\bibfnamefont {S.}~\bibnamefont {Ghaisari}}, \bibinfo {author}
  {\bibfnamefont {D.}~\bibnamefont {Spoddig}}, \bibinfo {author} {\bibfnamefont
  {D.}~\bibnamefont {Faivre}}, \bibinfo {author} {\bibfnamefont
  {R.}~\bibnamefont {Meckenstock}}, \bibinfo {author} {\bibfnamefont
  {M.}~\bibnamefont {Farle}},\ and\ \bibinfo {author} {\bibfnamefont
  {M.}~\bibnamefont {Winklhofer}},\ }\href
  {https://doi.org/10.1038/s41467-019-12219-0} {\bibfield  {journal} {\bibinfo
  {journal} {Nature Communications}\ }\textbf {\bibinfo {volume} {10}},\
  \bibinfo {pages} {4345} (\bibinfo {year} {2019})}\BibitemShut {NoStop}%
\bibitem [{\citenamefont {Rosenblatt}\ \emph {et~al.}(1982)\citenamefont
  {Rosenblatt}, \citenamefont {de~Araujo},\ and\ \citenamefont
  {Frankel}}]{Rosenblatt1982}%
  \BibitemOpen
  \bibfield  {author} {\bibinfo {author} {\bibfnamefont {C.}~\bibnamefont
  {Rosenblatt}}, \bibinfo {author} {\bibfnamefont {F.~F.~T.}\ \bibnamefont
  {de~Araujo}},\ and\ \bibinfo {author} {\bibfnamefont {R.~B.}\ \bibnamefont
  {Frankel}},\ }\href {https://doi.org/10.1063/1.330948} {\bibfield  {journal}
  {\bibinfo  {journal} {Journal of Applied Physics}\ }\textbf {\bibinfo
  {volume} {53}},\ \bibinfo {pages} {2727} (\bibinfo {year}
  {1982})}\BibitemShut {NoStop}%
\bibitem [{\citenamefont {Faivre}\ \emph {et~al.}(2010)\citenamefont {Faivre},
  \citenamefont {Fischer}, \citenamefont {Garcia-Rubio}, \citenamefont
  {Mastrogiacomo},\ and\ \citenamefont {Gehring}}]{Faivre2010}%
  \BibitemOpen
  \bibfield  {author} {\bibinfo {author} {\bibfnamefont {D.}~\bibnamefont
  {Faivre}}, \bibinfo {author} {\bibfnamefont {A.}~\bibnamefont {Fischer}},
  \bibinfo {author} {\bibfnamefont {I.}~\bibnamefont {Garcia-Rubio}}, \bibinfo
  {author} {\bibfnamefont {G.}~\bibnamefont {Mastrogiacomo}},\ and\ \bibinfo
  {author} {\bibfnamefont {A.}~\bibnamefont {Gehring}},\ }\href
  {https://doi.org/10.1016/j.bpj.2010.05.034} {\bibfield  {journal} {\bibinfo
  {journal} {Biophysical Journal}\ }\textbf {\bibinfo {volume} {99}},\ \bibinfo
  {pages} {1268} (\bibinfo {year} {2010})}\BibitemShut {NoStop}%
\bibitem [{\citenamefont {Fern{\'{a}}ndez-Castan{\'{e}}}\ \emph
  {et~al.}(2018)\citenamefont {Fern{\'{a}}ndez-Castan{\'{e}}}, \citenamefont
  {Li}, \citenamefont {Thomas},\ and\ \citenamefont {Overton}}]{Fernandez2018}%
  \BibitemOpen
  \bibfield  {author} {\bibinfo {author} {\bibfnamefont {A.}~\bibnamefont
  {Fern{\'{a}}ndez-Castan{\'{e}}}}, \bibinfo {author} {\bibfnamefont
  {H.}~\bibnamefont {Li}}, \bibinfo {author} {\bibfnamefont {O.~R.}\
  \bibnamefont {Thomas}},\ and\ \bibinfo {author} {\bibfnamefont {T.~W.}\
  \bibnamefont {Overton}},\ }\href {https://doi.org/10.1016/j.nbt.2018.05.1201}
  {\bibfield  {journal} {\bibinfo  {journal} {New Biotechnology}\ }\textbf
  {\bibinfo {volume} {46}},\ \bibinfo {pages} {22} (\bibinfo {year}
  {2018})}\BibitemShut {NoStop}%
\bibitem [{\citenamefont {Yang}\ \emph {et~al.}(2013)\citenamefont {Yang},
  \citenamefont {Li}, \citenamefont {Huang}, \citenamefont {Tang},
  \citenamefont {Jiang}, \citenamefont {Zhang},\ and\ \citenamefont
  {Li}}]{Yang2013a}%
  \BibitemOpen
  \bibfield  {author} {\bibinfo {author} {\bibfnamefont {J.}~\bibnamefont
  {Yang}}, \bibinfo {author} {\bibfnamefont {S.}~\bibnamefont {Li}}, \bibinfo
  {author} {\bibfnamefont {X.}~\bibnamefont {Huang}}, \bibinfo {author}
  {\bibfnamefont {T.}~\bibnamefont {Tang}}, \bibinfo {author} {\bibfnamefont
  {W.}~\bibnamefont {Jiang}}, \bibinfo {author} {\bibfnamefont
  {T.}~\bibnamefont {Zhang}},\ and\ \bibinfo {author} {\bibfnamefont
  {Y.}~\bibnamefont {Li}},\ }\href {https://doi.org/10.3389/fmicb.2013.00210}
  {\bibfield  {journal} {\bibinfo  {journal} {Frontiers in Microbiology}\
  }\textbf {\bibinfo {volume} {4}},\  (\bibinfo {year} {2013})}\BibitemShut
  {NoStop}%
\bibitem [{\citenamefont {Song}\ \emph {et~al.}(2014)\citenamefont {Song},
  \citenamefont {Zhao},\ and\ \citenamefont {Wu}}]{Song2014}%
  \BibitemOpen
  \bibfield  {author} {\bibinfo {author} {\bibfnamefont {T.}~\bibnamefont
  {Song}}, \bibinfo {author} {\bibfnamefont {L.}~\bibnamefont {Zhao}},\ and\
  \bibinfo {author} {\bibfnamefont {L.-F.}\ \bibnamefont {Wu}},\ }\href
  {https://doi.org/10.1109/TMAG.2014.2323953} {\bibfield  {journal} {\bibinfo
  {journal} {IEEE Transactions on Magnetics}\ }\textbf {\bibinfo {volume}
  {50}},\  (\bibinfo {year} {2014})}\BibitemShut {NoStop}%
\bibitem [{\citenamefont {Lef\`evre}\ \emph {et~al.}(2009)\citenamefont
  {Lef\`evre}, \citenamefont {Song}, \citenamefont {Yonnet},\ and\
  \citenamefont {Wu}}]{Lefevre2009}%
  \BibitemOpen
  \bibfield  {author} {\bibinfo {author} {\bibfnamefont {C.}~\bibnamefont
  {Lef\`evre}}, \bibinfo {author} {\bibfnamefont {T.}~\bibnamefont {Song}},
  \bibinfo {author} {\bibfnamefont {J.-P.}\ \bibnamefont {Yonnet}},\ and\
  \bibinfo {author} {\bibfnamefont {L.-F.}\ \bibnamefont {Wu}},\ }\href
  {https://doi.org/10.1128/AEM.00165-09} {\bibfield  {journal} {\bibinfo
  {journal} {Applied and Environmental Microbiology}\ }\textbf {\bibinfo
  {volume} {75}},\ \bibinfo {pages} {3835} (\bibinfo {year}
  {2009})}\BibitemShut {NoStop}%
\bibitem [{\citenamefont {Chen}\ \emph {et~al.}(2014)\citenamefont {Chen},
  \citenamefont {Chen}, \citenamefont {Yi}, \citenamefont {Chen}, \citenamefont
  {Wu},\ and\ \citenamefont {Song}}]{Chen2014}%
  \BibitemOpen
  \bibfield  {author} {\bibinfo {author} {\bibfnamefont {C.-Y.}\ \bibnamefont
  {Chen}}, \bibinfo {author} {\bibfnamefont {C.-F.}\ \bibnamefont {Chen}},
  \bibinfo {author} {\bibfnamefont {Y.}~\bibnamefont {Yi}}, \bibinfo {author}
  {\bibfnamefont {L.-J.}\ \bibnamefont {Chen}}, \bibinfo {author}
  {\bibfnamefont {L.-F.}\ \bibnamefont {Wu}},\ and\ \bibinfo {author}
  {\bibfnamefont {T.}~\bibnamefont {Song}},\ }\href
  {https://doi.org/10.1007/s10544-014-9880-2} {\bibfield  {journal} {\bibinfo
  {journal} {Biomedical Microdevices}\ }\textbf {\bibinfo {volume} {16}},\
  \bibinfo {pages} {761} (\bibinfo {year} {2014})}\BibitemShut {NoStop}%
\bibitem [{\citenamefont {Sch\"uler}\ \emph {et~al.}(1995)\citenamefont
  {Sch\"uler}, \citenamefont {Uhl},\ and\ \citenamefont
  {B\"auerlein}}]{Schuler1995}%
  \BibitemOpen
  \bibfield  {author} {\bibinfo {author} {\bibfnamefont {D.}~\bibnamefont
  {Sch\"uler}}, \bibinfo {author} {\bibfnamefont {R.}~\bibnamefont {Uhl}},\
  and\ \bibinfo {author} {\bibfnamefont {E.}~\bibnamefont {B\"auerlein}},\
  }\href {https://doi.org/10.1016/0378-1097(95)00300-T} {\bibfield  {journal}
  {\bibinfo  {journal} {FEMS Microbiology Letters}\ }\textbf {\bibinfo {volume}
  {132}},\ \bibinfo {pages} {139} (\bibinfo {year} {1995})}\BibitemShut
  {NoStop}%
\bibitem [{\citenamefont {Bennet}\ \emph {et~al.}(2016)\citenamefont {Bennet},
  \citenamefont {Gur}, \citenamefont {Yoon}, \citenamefont {Park},\ and\
  \citenamefont {Faivre}}]{Bennet2016}%
  \BibitemOpen
  \bibfield  {author} {\bibinfo {author} {\bibfnamefont {M.}~\bibnamefont
  {Bennet}}, \bibinfo {author} {\bibfnamefont {D.}~\bibnamefont {Gur}},
  \bibinfo {author} {\bibfnamefont {J.}~\bibnamefont {Yoon}}, \bibinfo {author}
  {\bibfnamefont {Y.}~\bibnamefont {Park}},\ and\ \bibinfo {author}
  {\bibfnamefont {D.}~\bibnamefont {Faivre}},\ }\href
  {https://doi.org/10.1002/adom.201600617} {\bibfield  {journal} {\bibinfo
  {journal} {Advanced Optical Materials}\ }\textbf {\bibinfo {volume} {5}},\
  \bibinfo {pages} {1600617} (\bibinfo {year} {2016})}\BibitemShut {NoStop}%
\bibitem [{\citenamefont {Katzmann}\ \emph {et~al.}(2013)\citenamefont
  {Katzmann}, \citenamefont {Eibauer}, \citenamefont {Lin}, \citenamefont
  {Pan}, \citenamefont {Plitzko},\ and\ \citenamefont
  {Schuler}}]{Katzmann2013}%
  \BibitemOpen
  \bibfield  {author} {\bibinfo {author} {\bibfnamefont {E.}~\bibnamefont
  {Katzmann}}, \bibinfo {author} {\bibfnamefont {M.}~\bibnamefont {Eibauer}},
  \bibinfo {author} {\bibfnamefont {W.}~\bibnamefont {Lin}}, \bibinfo {author}
  {\bibfnamefont {Y.}~\bibnamefont {Pan}}, \bibinfo {author} {\bibfnamefont
  {J.}~\bibnamefont {Plitzko}},\ and\ \bibinfo {author} {\bibfnamefont
  {D.}~\bibnamefont {Schuler}},\ }\href {https://doi.org/10.1128/AEM.02143-13}
  {\bibfield  {journal} {\bibinfo  {journal} {Applied and Environmental
  Microbiology}\ }\textbf {\bibinfo {volume} {79}},\ \bibinfo {pages} {7755}
  (\bibinfo {year} {2013})}\BibitemShut {NoStop}%
\bibitem [{\citenamefont {Myers}\ \emph {et~al.}(2013)\citenamefont {Myers},
  \citenamefont {Curtis},\ and\ \citenamefont {Curtis}}]{Myers2013}%
  \BibitemOpen
  \bibfield  {author} {\bibinfo {author} {\bibfnamefont {J.~A.}\ \bibnamefont
  {Myers}}, \bibinfo {author} {\bibfnamefont {B.~S.}\ \bibnamefont {Curtis}},\
  and\ \bibinfo {author} {\bibfnamefont {W.~R.}\ \bibnamefont {Curtis}},\
  }\href {https://doi.org/10.1186/2046-1682-6-4} {\bibfield  {journal}
  {\bibinfo  {journal} {{BMC} Biophysics}\ }\textbf {\bibinfo {volume} {6}},\
  \bibinfo {pages} {4} (\bibinfo {year} {2013})}\BibitemShut {NoStop}%
\bibitem [{\citenamefont {Zhao}\ \emph {et~al.}(2007)\citenamefont {Zhao},
  \citenamefont {Wu}, \citenamefont {Wu},\ and\ \citenamefont
  {Song}}]{Zhao2007}%
  \BibitemOpen
  \bibfield  {author} {\bibinfo {author} {\bibfnamefont {L.}~\bibnamefont
  {Zhao}}, \bibinfo {author} {\bibfnamefont {D.}~\bibnamefont {Wu}}, \bibinfo
  {author} {\bibfnamefont {L.-F.}\ \bibnamefont {Wu}},\ and\ \bibinfo {author}
  {\bibfnamefont {T.}~\bibnamefont {Song}},\ }\href
  {https://doi.org/10.1016/j.jbbm.2006.08.010} {\bibfield  {journal} {\bibinfo
  {journal} {Journal of Biochemical and Biophysical Methods}\ }\textbf
  {\bibinfo {volume} {70}},\ \bibinfo {pages} {377} (\bibinfo {year}
  {2007})}\BibitemShut {NoStop}%
\bibitem [{\citenamefont {Lang}\ and\ \citenamefont
  {Sch\"uler}(2008)}]{Lang2008}%
  \BibitemOpen
  \bibfield  {author} {\bibinfo {author} {\bibfnamefont {C.}~\bibnamefont
  {Lang}}\ and\ \bibinfo {author} {\bibfnamefont {D.}~\bibnamefont
  {Sch\"uler}},\ }\href {https://doi.org/10.1128/AEM.00231-08} {\bibfield
  {journal} {\bibinfo  {journal} {Applied and Environmental Microbiology}\
  }\textbf {\bibinfo {volume} {74}},\ \bibinfo {pages} {4944} (\bibinfo {year}
  {2008})}\BibitemShut {NoStop}%
\bibitem [{\citenamefont {Sch\"uler}\ and\ \citenamefont
  {Baeuerlein}(1998)}]{Schuler1998}%
  \BibitemOpen
  \bibfield  {author} {\bibinfo {author} {\bibfnamefont {D.}~\bibnamefont
  {Sch\"uler}}\ and\ \bibinfo {author} {\bibfnamefont {E.}~\bibnamefont
  {Baeuerlein}},\ }\href
  {https://www.scopus.com/inward/record.uri?eid=2-s2.0-0031984106&partnerID=40&md5=8f35d2ee1cd7d31775b66ea6b39de064}
  {\bibfield  {journal} {\bibinfo  {journal} {Journal of Bacteriology}\
  }\textbf {\bibinfo {volume} {180}},\ \bibinfo {pages} {159} (\bibinfo {year}
  {1998})}\BibitemShut {NoStop}%
\bibitem [{\citenamefont {Heyen}\ and\ \citenamefont
  {Sch\"uler}(2003)}]{Heyen2003}%
  \BibitemOpen
  \bibfield  {author} {\bibinfo {author} {\bibfnamefont {U.}~\bibnamefont
  {Heyen}}\ and\ \bibinfo {author} {\bibfnamefont {D.}~\bibnamefont
  {Sch\"uler}},\ }\href
  {https://www.scopus.com/inward/record.uri?eid=2-s2.0-0038236515&partnerID=40&md5=5f6f9b12fa1d919ab34d9865e5f9e55a}
  {\bibfield  {journal} {\bibinfo  {journal} {Applied Microbiology and
  Biotechnology}\ }\textbf {\bibinfo {volume} {61}},\ \bibinfo {pages} {536}
  (\bibinfo {year} {2003})}\BibitemShut {NoStop}%
\bibitem [{\citenamefont {Esquivel}\ and\ \citenamefont {Lins~de
  Barros}(1986)}]{Esquivel1986}%
  \BibitemOpen
  \bibfield  {author} {\bibinfo {author} {\bibfnamefont {D.}~\bibnamefont
  {Esquivel}}\ and\ \bibinfo {author} {\bibfnamefont {H.}~\bibnamefont {Lins~de
  Barros}},\ }\href
  {https://www.scopus.com/inward/record.uri?eid=2-s2.0-0022620465&partnerID=40&md5=c20241b29698f8791909f768825f8fdf}
  {\bibfield  {journal} {\bibinfo  {journal} {Journal of Experimental Biology}\
  }\textbf {\bibinfo {volume} {121}},\ \bibinfo {pages} {153} (\bibinfo {year}
  {1986})}\BibitemShut {NoStop}%
\bibitem [{\citenamefont {Steinberger}\ \emph {et~al.}(1994)\citenamefont
  {Steinberger}, \citenamefont {Petersen}, \citenamefont {Petermann},\ and\
  \citenamefont {Weiss}}]{Steinberger1994}%
  \BibitemOpen
  \bibfield  {author} {\bibinfo {author} {\bibfnamefont {B.}~\bibnamefont
  {Steinberger}}, \bibinfo {author} {\bibfnamefont {N.}~\bibnamefont
  {Petersen}}, \bibinfo {author} {\bibfnamefont {H.}~\bibnamefont
  {Petermann}},\ and\ \bibinfo {author} {\bibfnamefont {D.~G.}\ \bibnamefont
  {Weiss}},\ }\href {https://doi.org/10.1017/s0022112094001904} {\bibfield
  {journal} {\bibinfo  {journal} {Journal of Fluid Mechanics}\ }\textbf
  {\bibinfo {volume} {273}},\ \bibinfo {pages} {189} (\bibinfo {year}
  {1994})}\BibitemShut {NoStop}%
\bibitem [{\citenamefont {Zahn}\ \emph {et~al.}(2017)\citenamefont {Zahn},
  \citenamefont {Keller}, \citenamefont {Toro-Nahuelpan}, \citenamefont
  {Dorscht}, \citenamefont {Gross}, \citenamefont {Laumann}, \citenamefont
  {Gekle}, \citenamefont {Zimmermann}, \citenamefont {Sch\"uler},\ and\
  \citenamefont {Kress}}]{Zahn2017}%
  \BibitemOpen
  \bibfield  {author} {\bibinfo {author} {\bibfnamefont {C.}~\bibnamefont
  {Zahn}}, \bibinfo {author} {\bibfnamefont {S.}~\bibnamefont {Keller}},
  \bibinfo {author} {\bibfnamefont {M.}~\bibnamefont {Toro-Nahuelpan}},
  \bibinfo {author} {\bibfnamefont {P.}~\bibnamefont {Dorscht}}, \bibinfo
  {author} {\bibfnamefont {W.}~\bibnamefont {Gross}}, \bibinfo {author}
  {\bibfnamefont {M.}~\bibnamefont {Laumann}}, \bibinfo {author} {\bibfnamefont
  {S.}~\bibnamefont {Gekle}}, \bibinfo {author} {\bibfnamefont
  {W.}~\bibnamefont {Zimmermann}}, \bibinfo {author} {\bibfnamefont
  {D.}~\bibnamefont {Sch\"uler}},\ and\ \bibinfo {author} {\bibfnamefont
  {H.}~\bibnamefont {Kress}},\ }\href
  {https://doi.org/10.1038/s41598-017-03756-z} {\bibfield  {journal} {\bibinfo
  {journal} {Scientific Reports}\ }\textbf {\bibinfo {volume} {7}},\  (\bibinfo
  {year} {2017})}\BibitemShut {NoStop}%
\bibitem [{\citenamefont {Erglis}\ \emph {et~al.}(2007)\citenamefont {Erglis},
  \citenamefont {Wen}, \citenamefont {Ose}, \citenamefont {Zeltins},
  \citenamefont {Sharipo}, \citenamefont {Janmey},\ and\ \citenamefont
  {Cebers}}]{Erglis2007}%
  \BibitemOpen
  \bibfield  {author} {\bibinfo {author} {\bibfnamefont {K.}~\bibnamefont
  {Erglis}}, \bibinfo {author} {\bibfnamefont {Q.}~\bibnamefont {Wen}},
  \bibinfo {author} {\bibfnamefont {V.}~\bibnamefont {Ose}}, \bibinfo {author}
  {\bibfnamefont {A.}~\bibnamefont {Zeltins}}, \bibinfo {author} {\bibfnamefont
  {A.}~\bibnamefont {Sharipo}}, \bibinfo {author} {\bibfnamefont {P.~A.}\
  \bibnamefont {Janmey}},\ and\ \bibinfo {author} {\bibfnamefont
  {A.}~\bibnamefont {Cebers}},\ }\href
  {https://doi.org/10.1529/biophysj.107.107474} {\bibfield  {journal} {\bibinfo
   {journal} {Biophysical Journal}\ }\textbf {\bibinfo {volume} {93}},\
  \bibinfo {pages} {1402} (\bibinfo {year} {2007})}\BibitemShut {NoStop}%
\bibitem [{\citenamefont {Pan}\ \emph {et~al.}(2009)\citenamefont {Pan},
  \citenamefont {Lin}, \citenamefont {Li}, \citenamefont {Wu}, \citenamefont
  {Tian}, \citenamefont {Deng}, \citenamefont {Liu}, \citenamefont {Zhu},
  \citenamefont {Winklhofer},\ and\ \citenamefont {Petersen}}]{Pan2009}%
  \BibitemOpen
  \bibfield  {author} {\bibinfo {author} {\bibfnamefont {Y.}~\bibnamefont
  {Pan}}, \bibinfo {author} {\bibfnamefont {W.}~\bibnamefont {Lin}}, \bibinfo
  {author} {\bibfnamefont {J.}~\bibnamefont {Li}}, \bibinfo {author}
  {\bibfnamefont {W.}~\bibnamefont {Wu}}, \bibinfo {author} {\bibfnamefont
  {L.}~\bibnamefont {Tian}}, \bibinfo {author} {\bibfnamefont {C.}~\bibnamefont
  {Deng}}, \bibinfo {author} {\bibfnamefont {Q.}~\bibnamefont {Liu}}, \bibinfo
  {author} {\bibfnamefont {R.}~\bibnamefont {Zhu}}, \bibinfo {author}
  {\bibfnamefont {M.}~\bibnamefont {Winklhofer}},\ and\ \bibinfo {author}
  {\bibfnamefont {N.}~\bibnamefont {Petersen}},\ }\href
  {https://doi.org/10.1016/j.bpj.2009.06.012} {\bibfield  {journal} {\bibinfo
  {journal} {Biophysical Journal}\ }\textbf {\bibinfo {volume} {97}},\ \bibinfo
  {pages} {986} (\bibinfo {year} {2009})}\BibitemShut {NoStop}%
\bibitem [{\citenamefont {Lane}\ \emph {et~al.}(2001)\citenamefont {Lane},
  \citenamefont {Youngquist}, \citenamefont {Immer},\ and\ \citenamefont
  {Simpson}}]{Lane2001}%
  \BibitemOpen
  \bibfield  {author} {\bibinfo {author} {\bibfnamefont {J.~E.}\ \bibnamefont
  {Lane}}, \bibinfo {author} {\bibfnamefont {R.~C.}\ \bibnamefont
  {Youngquist}}, \bibinfo {author} {\bibfnamefont {C.~D.}\ \bibnamefont
  {Immer}},\ and\ \bibinfo {author} {\bibfnamefont {J.~C.}\ \bibnamefont
  {Simpson}},\ }\href@noop {} {\bibfield  {journal} {\bibinfo  {journal}
  {National Aeronautics and Space Administration, Washington}\ ,\ \bibinfo
  {pages} {217918}} (\bibinfo {year} {2001})}\BibitemShut {NoStop}%
\bibitem [{\citenamefont {Hergert}(2014)}]{Hergert2014}%
  \BibitemOpen
  \bibfield  {author} {\bibinfo {author} {\bibfnamefont {E.}~\bibnamefont
  {Hergert}},\ }\href
  {https://hub.hamamatsu.com/us/en/technical-notes/detector-selection/guide-to-detector-selection.html}
  {\emph {\bibinfo {title} {Guide to Detector Selection}}},\ \bibinfo {type}
  {Tech. Rep.}\ (\bibinfo  {institution} {Hamamatsu Cooperation},\ \bibinfo
  {year} {2014})\BibitemShut {NoStop}%
\bibitem [{\citenamefont {Pfeiffer}\ and\ \citenamefont
  {Schüler}(2020)}]{Pfeiffer2020}%
  \BibitemOpen
  \bibfield  {author} {\bibinfo {author} {\bibfnamefont {D.}~\bibnamefont
  {Pfeiffer}}\ and\ \bibinfo {author} {\bibfnamefont {D.}~\bibnamefont
  {Schüler}},\ }\bibfield  {journal} {\bibinfo  {journal} {Applied and
  Environmental Microbiology}\ }\textbf {\bibinfo {volume} {86}},\ \href
  {https://doi.org/10.1128/aem.01976-19} {10.1128/aem.01976-19} (\bibinfo
  {year} {2020})\BibitemShut {NoStop}%
\bibitem [{\citenamefont {Popp}\ \emph {et~al.}(2014)\citenamefont {Popp},
  \citenamefont {Armitage},\ and\ \citenamefont {Sch{\"u}ler}}]{Popp2014}%
  \BibitemOpen
  \bibfield  {author} {\bibinfo {author} {\bibfnamefont {F.}~\bibnamefont
  {Popp}}, \bibinfo {author} {\bibfnamefont {J.}~\bibnamefont {Armitage}},\
  and\ \bibinfo {author} {\bibfnamefont {D.}~\bibnamefont {Sch{\"u}ler}},\
  }\href {https://doi.org/10.1038/ncomms6398} {\bibfield  {journal} {\bibinfo
  {journal} {Nature Communications}\ }\textbf {\bibinfo {volume} {5}},\
  (\bibinfo {year} {2014})}\BibitemShut {NoStop}%
\bibitem [{\citenamefont {Staniland}\ \emph {et~al.}(2010)\citenamefont
  {Staniland}, \citenamefont {Moisescu},\ and\ \citenamefont
  {Benning}}]{Staniland2010}%
  \BibitemOpen
  \bibfield  {author} {\bibinfo {author} {\bibfnamefont {S.~S.}\ \bibnamefont
  {Staniland}}, \bibinfo {author} {\bibfnamefont {C.}~\bibnamefont
  {Moisescu}},\ and\ \bibinfo {author} {\bibfnamefont {L.~G.}\ \bibnamefont
  {Benning}},\ }\href {https://doi.org/10.1002/jobm.200900408} {\bibfield
  {journal} {\bibinfo  {journal} {Journal of Basic Microbiology}\ }\textbf
  {\bibinfo {volume} {50}},\ \bibinfo {pages} {392} (\bibinfo {year}
  {2010})}\BibitemShut {NoStop}%
\bibitem [{\citenamefont {Yang}\ \emph {et~al.}(2001)\citenamefont {Yang},
  \citenamefont {Takeyama}, \citenamefont {Tanaka}, \citenamefont {Hasegawa},\
  and\ \citenamefont {Matsunaga}}]{Yang2001b}%
  \BibitemOpen
  \bibfield  {author} {\bibinfo {author} {\bibfnamefont {C.-D.}\ \bibnamefont
  {Yang}}, \bibinfo {author} {\bibfnamefont {H.}~\bibnamefont {Takeyama}},
  \bibinfo {author} {\bibfnamefont {T.}~\bibnamefont {Tanaka}}, \bibinfo
  {author} {\bibfnamefont {A.}~\bibnamefont {Hasegawa}},\ and\ \bibinfo
  {author} {\bibfnamefont {T.}~\bibnamefont {Matsunaga}},\ }\href
  {https://doi.org/10.1385/abab:91-93:1-9:155} {\bibfield  {journal} {\bibinfo
  {journal} {Applied Biochemistry and Biotechnology}\ }\textbf {\bibinfo
  {volume} {91-93}},\ \bibinfo {pages} {155} (\bibinfo {year}
  {2001})}\BibitemShut {NoStop}%
\bibitem [{\citenamefont {Pichel}(2018)}]{Pichel2018a}%
  \BibitemOpen
  \bibfield  {author} {\bibinfo {author} {\bibfnamefont {M.}~\bibnamefont
  {Pichel}},\ }\emph {\bibinfo {title} {The behavior of magnetotacticbacteria
  in changing magnetic fields}},\ \href
  {https://doi.org/10.3990/1.9789036545006} {Ph.D. thesis},\ \bibinfo  {school}
  {University of Twente} (\bibinfo {year} {2018})\BibitemShut {NoStop}%
\bibitem [{\citenamefont {Bazylinski}\ \emph {et~al.}(2013)\citenamefont
  {Bazylinski}, \citenamefont {Williams}, \citenamefont {Lef{\`{e}}vre},
  \citenamefont {Berg}, \citenamefont {Zhang}, \citenamefont {Bowser},
  \citenamefont {Dean},\ and\ \citenamefont {Beveridge}}]{Bazylinski2013}%
  \BibitemOpen
  \bibfield  {author} {\bibinfo {author} {\bibfnamefont {D.~A.}\ \bibnamefont
  {Bazylinski}}, \bibinfo {author} {\bibfnamefont {T.~J.}\ \bibnamefont
  {Williams}}, \bibinfo {author} {\bibfnamefont {C.~T.}\ \bibnamefont
  {Lef{\`{e}}vre}}, \bibinfo {author} {\bibfnamefont {R.~J.}\ \bibnamefont
  {Berg}}, \bibinfo {author} {\bibfnamefont {C.~L.}\ \bibnamefont {Zhang}},
  \bibinfo {author} {\bibfnamefont {S.~S.}\ \bibnamefont {Bowser}}, \bibinfo
  {author} {\bibfnamefont {A.~J.}\ \bibnamefont {Dean}},\ and\ \bibinfo
  {author} {\bibfnamefont {T.~J.}\ \bibnamefont {Beveridge}},\ }\href
  {https://doi.org/10.1099/ijs.0.038927-0} {\bibfield  {journal} {\bibinfo
  {journal} {International Journal of Systematic and Evolutionary
  Microbiology}\ }\textbf {\bibinfo {volume} {63}},\ \bibinfo {pages} {801}
  (\bibinfo {year} {2013})}\BibitemShut {NoStop}%
\bibitem [{\citenamefont {Lef{\`{e}}vre}\ \emph {et~al.}(2016)\citenamefont
  {Lef{\`{e}}vre}, \citenamefont {Howse}, \citenamefont {Schmidt},
  \citenamefont {Sabaty}, \citenamefont {Menguy}, \citenamefont {Luther},\ and\
  \citenamefont {Bazylinski}}]{Lefevre2016}%
  \BibitemOpen
  \bibfield  {author} {\bibinfo {author} {\bibfnamefont {C.~T.}\ \bibnamefont
  {Lef{\`{e}}vre}}, \bibinfo {author} {\bibfnamefont {P.~A.}\ \bibnamefont
  {Howse}}, \bibinfo {author} {\bibfnamefont {M.~L.}\ \bibnamefont {Schmidt}},
  \bibinfo {author} {\bibfnamefont {M.}~\bibnamefont {Sabaty}}, \bibinfo
  {author} {\bibfnamefont {N.}~\bibnamefont {Menguy}}, \bibinfo {author}
  {\bibfnamefont {G.~W.}\ \bibnamefont {Luther}},\ and\ \bibinfo {author}
  {\bibfnamefont {D.~A.}\ \bibnamefont {Bazylinski}},\ }\href
  {https://doi.org/10.1111/1758-2229.12479} {\bibfield  {journal} {\bibinfo
  {journal} {Environmental Microbiology Reports}\ }\textbf {\bibinfo {volume}
  {8}},\ \bibinfo {pages} {1003} (\bibinfo {year} {2016})}\BibitemShut
  {NoStop}%
\bibitem [{\citenamefont {Ahlers}\ \emph {et~al.}(2022)\citenamefont {Ahlers},
  \citenamefont {Block}, \citenamefont {Winklhofer},\ and\ \citenamefont
  {Greschner}}]{Ahlers2022}%
  \BibitemOpen
  \bibfield  {author} {\bibinfo {author} {\bibfnamefont {M.~T.}\ \bibnamefont
  {Ahlers}}, \bibinfo {author} {\bibfnamefont {C.~T.}\ \bibnamefont {Block}},
  \bibinfo {author} {\bibfnamefont {M.}~\bibnamefont {Winklhofer}},\ and\
  \bibinfo {author} {\bibfnamefont {M.}~\bibnamefont {Greschner}},\ }\href
  {https://doi.org/10.1371/journal.pone.0271765} {\bibfield  {journal}
  {\bibinfo  {journal} {{PLOS} {ONE}}\ }\textbf {\bibinfo {volume} {17}},\
  \bibinfo {pages} {e0271765} (\bibinfo {year} {2022})}\BibitemShut {NoStop}%
\bibitem [{\citenamefont {Song}\ \emph {et~al.}(2021)\citenamefont {Song},
  \citenamefont {Yoon}, \citenamefont {Jeong}, \citenamefont {Jung},
  \citenamefont {Abelmann},\ and\ \citenamefont {Park}}]{Song2021}%
  \BibitemOpen
  \bibfield  {author} {\bibinfo {author} {\bibfnamefont {S.-H.}\ \bibnamefont
  {Song}}, \bibinfo {author} {\bibfnamefont {J.}~\bibnamefont {Yoon}}, \bibinfo
  {author} {\bibfnamefont {Y.}~\bibnamefont {Jeong}}, \bibinfo {author}
  {\bibfnamefont {Y.-G.}\ \bibnamefont {Jung}}, \bibinfo {author}
  {\bibfnamefont {L.}~\bibnamefont {Abelmann}},\ and\ \bibinfo {author}
  {\bibfnamefont {W.}~\bibnamefont {Park}},\ }\href
  {https://doi.org/10.1016/j.jmmm.2021.168341} {\bibfield  {journal} {\bibinfo
  {journal} {Journal of Magnetism and Magnetic Materials}\ }\textbf {\bibinfo
  {volume} {539}},\ \bibinfo {pages} {168341} (\bibinfo {year}
  {2021})}\BibitemShut {NoStop}%
\bibitem [{\citenamefont {L{\"o}thman}\ \emph {et~al.}(2018)\citenamefont
  {L{\"o}thman}, \citenamefont {Hageman}, \citenamefont {Hendrix},
  \citenamefont {Keizer}, \citenamefont {{van Wolferen}}, \citenamefont {Ma},
  \citenamefont {{van de Loosdrecht}}, \citenamefont {{ten Haken}},
  \citenamefont {Bolhuis},\ and\ \citenamefont {Abelmann}}]{Loethman2018a}%
  \BibitemOpen
  \bibfield  {author} {\bibinfo {author} {\bibfnamefont {P.}~\bibnamefont
  {L{\"o}thman}}, \bibinfo {author} {\bibfnamefont {T.}~\bibnamefont
  {Hageman}}, \bibinfo {author} {\bibfnamefont {J.}~\bibnamefont {Hendrix}},
  \bibinfo {author} {\bibfnamefont {H.}~\bibnamefont {Keizer}}, \bibinfo
  {author} {\bibfnamefont {H.}~\bibnamefont {{van Wolferen}}}, \bibinfo
  {author} {\bibfnamefont {K.}~\bibnamefont {Ma}}, \bibinfo {author}
  {\bibfnamefont {M.}~\bibnamefont {{van de Loosdrecht}}}, \bibinfo {author}
  {\bibfnamefont {B.}~\bibnamefont {{ten Haken}}}, \bibinfo {author}
  {\bibfnamefont {T.}~\bibnamefont {Bolhuis}},\ and\ \bibinfo {author}
  {\bibfnamefont {L.}~\bibnamefont {Abelmann}},\ }\href@noop {} {\bibfield
  {journal} {\bibinfo  {journal} {8th International Workshop on Magnetic
  Particle Imaging}\ } (\bibinfo {year} {2018})}\BibitemShut {NoStop}%
\bibitem [{\citenamefont {Gao}\ \emph {et~al.}(2013)\citenamefont {Gao},
  \citenamefont {van Reenen}, \citenamefont {Hulsen}, \citenamefont {de~Jong},
  \citenamefont {Prins},\ and\ \citenamefont {den Toonder}}]{Gao2013}%
  \BibitemOpen
  \bibfield  {author} {\bibinfo {author} {\bibfnamefont {Y.}~\bibnamefont
  {Gao}}, \bibinfo {author} {\bibfnamefont {A.}~\bibnamefont {van Reenen}},
  \bibinfo {author} {\bibfnamefont {M.~A.}\ \bibnamefont {Hulsen}}, \bibinfo
  {author} {\bibfnamefont {A.~M.}\ \bibnamefont {de~Jong}}, \bibinfo {author}
  {\bibfnamefont {M.~W.~J.}\ \bibnamefont {Prins}},\ and\ \bibinfo {author}
  {\bibfnamefont {J.~M.~J.}\ \bibnamefont {den Toonder}},\ }\href
  {https://doi.org/10.1039/c3lc41229f} {\bibfield  {journal} {\bibinfo
  {journal} {Lab on a Chip}\ }\textbf {\bibinfo {volume} {13}},\ \bibinfo
  {pages} {1394} (\bibinfo {year} {2013})}\BibitemShut {NoStop}%
\end{thebibliography}%

\end{document}